\documentclass[fleqn,10pt]{wlscirep}
\usepackage[utf8]{inputenc}
\usepackage[T1]{fontenc}
\usepackage{microtype}
\usepackage{graphicx}
\usepackage[export]{adjustbox}
\usepackage{subfigure}
\usepackage{booktabs} 
\usepackage{hyperref}

\RequirePackage{algorithm,algpseudocode}
\usepackage{amsmath}
\usepackage{amssymb}
\usepackage{mathtools}
\usepackage{amsthm}
\usepackage[capitalize,noabbrev]{cleveref}
\theoremstyle{plain}
\newtheorem{theorem}{Theorem}
\newtheorem{proposition}{Proposition}
\newtheorem{lemma}{Lemma}
\newtheorem{corollary}{Corollary}
\newtheorem{definition}{Definition}
\newtheorem{assumption}{Assumption}
\newtheorem{remark}{Remark}
\newtheorem{Ex}{Example}
\usepackage[textsize=tiny]{todonotes}
\title{Degree-corrected distribution-free model for community detection in weighted networks}
\author[*]{Huan Qing}
\affil[*]{China University of Mining and Technology, School of Mathematics, Xuzhou, 221116, P.R. China}
\affil[*]{qinghuan@cumt.edu.cn}

\begin{abstract}
A degree-corrected distribution-free model is proposed for weighted social networks with latent structural information. The model extends the previous distribution-free models by considering variation in node degree to fit real-world weighted networks, and it also extends the classical degree-corrected stochastic block model from un-weighted network to weighted network. We design an algorithm based on the idea of spectral clustering to fit the model. Theoretical framework on consistent estimation for the algorithm is developed under the model. Theoretical results when edge weights are generated from different distributions are analyzed. We also propose a general modularity as an extension of Newman's modularity from un-weighted network to weighted network. Using experiments with simulated and real-world networks, we show that our method significantly outperforms the uncorrected one, and the general modularity is effective.
\end{abstract}
\begin{document}

\flushbottom
\maketitle
%
%
\thispagestyle{empty}
\section*{Introduction}
Network data analysis is an important research topic in a range of scientific disciplines in recent years, particularly in the biological science, social science, physics and computer science. Many researchers aim at analyzing these networks by developing models, quantitative tools and theoretical framework to have a deeper understanding of the underlying structural information. A problem in network science that is of major interest is ``community detection''. The Stochastic Blockmodels (SBM) \cite{SBM} is a classic model to model un-weighted networks for community detection.  In SBM, every node in the same community shares the same expectation degree, which is unrealistic for real-world networks since nodes degrees vary in most real-world networks. To overcome this limitation of SBM, the popular model Degree Corrected Stochastic Blockmodels (DCSBM) proposed in \cite{DCSBM} considers node heterogeneity to extend SBM by allowing that nodes in the same community can have various expectation degrees. Many community detection methods and theoretical studies have been developed under SBM and DCSBM, to name a few, \cite{abbe2015community,cai2015robust,abbe2016exact,chen2018convexified,amini2018on,su2020strong}, and references therein.

However, most works built under SBM and DCSBM require the elements of adjacency matrix of the network to follow Bernoulli distribution, which limits the network to being un-weighted. Modeling and designing methods to quantitatively detecting latent structural information for weighted networks are interesting topics. Recent years, some Weighted Stochastic Blockmodels (WSBM) have been developed for weighted networks, to name a few,  \cite{aicher2015learning, jog2015information,ahn2018hypergraph, palowitch2018significance,peixoto2018nonparametric,xu2020optimal,ng2021weighted}. However, though these models for weighted networks are attractive, they always require all elements of connectivity  matrix to be nonnegative or all elements of adjacency matrix must follow some specific distributions as found in \cite{qing2021DFM}. Furthermore, spectral clustering is widely used to study the structure of networks under SBM and DCSBM, for example, \cite{rohe2011spectral,RSC, SCORE,lei2015consistency,sengupta2015spectral,joseph2016impact}. Another limitation of the above WSBMs is,  it is challenging to develop methods by taking the advantage of the spectral clustering idea under these WSBMs for their complex forms or strict constraint on edge distribution. To overcome limitations of these weighted models, \cite{qing2021DFM} proposes a Distribution-Free Models (DFM) which has no requirement on the distribution of adjacency matrix's elements and allows developing methods to fit the model by taking the advantage of spectral clustering. DFM can be seen as a direct extension of SBM, and nodes within the same community under DFM shares same expectation degrees, which is unrealistic for empirical networks with various nodes degrees.

In this paper, we develop a model called Degree-Corrected Distribution-Free Model (DCDFM) as an extension of DFM by considering node heterogeneity. We extend the previous results in the following ways:

(a) DCDFM models weighted networks by allowing nodes within the same community to have different expectation degrees. Though the WSBM developed in \cite{palowitch2018significance} also considers node heterogeneity, it requires all elements of connectivity matrix to be nonnegative, and fitting it by spectral clustering is challenging. Our DCDFM inherits the advantages of DFM such that it has no constraint on distribution of adjacency matrix, allows connectivity matrix to have negative entries, and allows applying the idea of spectral clustering to fit it. Meanwhile, as an extension of DFM, similar as the relationship between SBM and DCSBM, nodes within the same community can have different expectation degrees under our DCDFM, and this ensures that DCDFM can model real-world weighted networks in which nodes have various degrees.

(b) To fit DCDFM, an efficient spectral clustering algorithm called nDFA is designed. We build theoretical framework on consistent estimation for the proposed algorithm under DCDFM. Benefited from the distribution-free property of DCDFM, our theoretical results under DCDFM are general. Especially, when DCDFM reduces to DFM, our theoretical results are consistent with those under DFM. When DCDFM degenerates to DCSBM, our results also match classical results under DCSBM. Numerical results of both simulated and real-world networks show the advantage of introducing node heterogeneity to model weighted networks.

(c) To measure performances of different methods on real-world weighted network with unknown information on nodes labels, we propose a general modularity as an extension of classical Newman's modularity \cite{newman2006modularity}. For weighted network in which all edge weights are nonnegative, the general modularity is exactly the Newman's modularity. For weighted network in which some edge weights are negative, the general modularity considers negative edge weights. Numerical results on simulated network generated under DCDFM for different distributions, and empirical un-weighted and weighted networks with known ground-truth nodes labels support the effectiveness of the general modularity. By using two community-oriented topological measures introduced in \cite{orman2012comparative}, we find that the modularity is effective and our nDFA returns reasonable community partition for real-world weighted networks with unknown ground-truth nodes labels.

\textbf{\textit{Notations.}}
We take the following general notations in this paper. For any positive integer $m$, let $[m]:= \{1,2,\ldots,m\}$ and $I_{m}$ be the $m\times m$ identity matrix. For a vector $x$, $\|x\|_{q}$ denotes its $l_{q}$-norm. $M'$ is the transpose of the matrix $M$, and $\|M\|$ denotes the spectral norm, $\|M\|_{F}$ denotes the Frobenius norm, and $\|M\|_{0}$ denotes the $l_{0}$ norm by counting the number of nonzero entries in $M$. For convenience, when we say ``leading eigenvalues'' or ``leading eigenvectors'', we are comparing the \emph{magnitudes} of the eigenvalues and their respective eigenvectors with unit-norm. Let $\lambda_{k}(M)$ be the $k$-th leading eigenvalue of matrix $M$. $M(i,:)$ and $M(:,j)$ denote the $i$-th row and the $j$-th column of matrix $M$, respectively. $M(S,:)$ denotes the rows in the index sets $S$ of matrix $M$. $\mathrm{rank}(M)$ denotes the rank of matrix $M$.
\section*{Degree-Corrected Distribution-Free Model}\label{modelDCDFM}
Let $\mathcal{N}$ be an undirected weighted network with $n$ nodes. Let $A$ be the $n\times n$ symmetric adjacency matrix of $\mathcal{N}$, and $A(i,j)$ denotes the weight between node $i$ and node $j$ for all node pairs. Since we consider weighted network, $A(i,j)$ is finite real values, and it can even be negative for $i,j\in[n]$. Throughout this article, we assume that in network $\mathcal{N}$, all nodes belong to
\begin{align}\label{DefinSC}
K~\mathrm{nonoverlapping~ communities~}\{\mathcal{C}_{k}\}^{K}_{k=1}.
\end{align}
Let $\ell$ be an $n\times 1$ vector such that $\ell(i)=k$ if node $i$ belongs to community $k$ for $i\in[n], k\in[K]$. For convenience, let $Z\in \{0,1\}^{n\times K}$ be the membership matrix such that for $i\in[n]$
\begin{align}\label{DefineSPMF}
\mathrm{rank}(Z)=K, \mathrm{and~}Z(i,\ell(i))=1, \|Z(i,:)\|_{1}=1.
\end{align}
$\mathrm{rank}(Z)=K$ means that each community $\mathcal{C}_{k}$ has at least one node for $k\in[K]$. $\mathrm{and~}Z(i,\ell(i))=1, \|Z(i,:)\|_{1}=1$ mean that $Z(i,k)=1$ if $k=\ell(i)$ and $Z(i,k)=0$ if $k\neq \ell(i)$, for $i\in[n],k\in[K]$, i.e., each node only belongs to one of the $K$ communities.

Let $n_{k}=|i\in[n]:\ell(i)=k|$ be the size of community $k$ for $k\in[K]$. Set $n_{\mathrm{max}}=\mathrm{max}_{k\in[K]}n_{k}, n_{\mathrm{min}}=\mathrm{min}_{k\in[K]}n_{k}$. Let the connectivity matrix $P\in \mathbb{R}^{K\times K}$ satisfy
\begin{align}\label{definP}
 P=P', \mathrm{rank}(P)=K, \mathrm{and~}|P_{\mathrm{max}}|=1,
\end{align}
where $|P_{\mathrm{max}}|=\mathrm{max}_{k,l\in[K]}|P(k,l)|$. Eq (\ref{definP}) means that $P$ is a full rank symmetric matrix, and we set the maximum absolute value of $P$'s entries as 1 mainly for convenience. Meanwhile, it should be emphasized that Eq (\ref{definP}) allows $P$ to have negative elements. Unless specified, $K$ is assumed to be known in this paper.

Let $\theta$ be an $n\times 1$ vector such that $\theta(i)$ is the node heterogeneity parameter (also known as degree heterogeneity) of node $i$, for $i\in[n]$. Let $\Theta$ be an $n\times n$ diagonal matrix whose $i$-th diagonal element is $\theta(i)$. For convenience, set $\theta_{\mathrm{max}}=\mathrm{max}_{i\in[n]}\theta(i)$, and $\theta_{\mathrm{min}}=\mathrm{min}_{i\in[n]}\theta(i)$. Since all entries of $\theta$ are node heterogeneities, we have
\begin{align}\label{DefinTheta}
\theta_{\mathrm{min}}>0.
\end{align}
For \emph{arbitrary distribution} $\mathcal{F}$, and all pairs of $(i,j)$ with $i,j\in[n]$, our model assumes that $A(i,j)$ are independent random variables generated according to $\mathcal{F}$ with expectation
\begin{align}\label{DefinMM}
\mathbb{E}[A(i,j)]=\Omega(i,j), \mathrm{where~}\Omega:=\Theta ZPZ'\Theta.
\end{align}
Eq (\ref{DefinMM}) means that we only assume all elements of $A$ are independent random variables and $\mathbb{E}[A]=\Theta ZPZ'\Theta$ without any prior knowledge on specific distribution of $A(i,j)$ for $i,j\in[n]$ since distribution $\mathcal{F}$ can be arbitrary. The rationality of our assumption on the arbitrariness of distribution $\mathcal{F}$ comes from the fact that we can generate a random number $A(i,j)$ from distribution $\mathcal{F}$ with expectation $\Omega(i,j)$. So, instead of fixing $\mathcal{F}$ to be a special distribution, $A$ is allowed to be generated from any distribution $\mathcal{F}$ as long as the block structure in Eq (\ref{DefinMM}) holds under DCDFM.
\begin{definition}
Call model (\ref{DefinSC})-(\ref{DefinMM}) the Degree-Corrected Distribution-Free Model (DCDFM), and denote it by $DCDFM_{n}(K,P,Z,\Theta)$.
\end{definition}
Our model DCDFM also inherits the distribution-free property of DFM by Eq (\ref{DefinMM}). Remarks on understanding $Z,P,\theta$, node degree, network connectivity and self-connected nodes are provided below.
\begin{remark}
For DCDFM, node label $\ell$ is defined by $Z$ satisfying Eq (\ref{DefineSPMF}). Actually, when defining $\ell$, we can reduce requirement on $Z$ such that $Z\in\mathbb{R}^{n\times K}$ has $K$ distinct rows. For such case, to define node memberships, let two distinct nodes $i$ and $j$ be in the same cluster as long as $Z(i,:)=Z(j,:)$. All theoretical results in this paper remains the same  under DCDFM constructed for such $Z$. In this paper, we consider $Z$ satisfying Eq (\ref{DefineSPMF}) mainly for convenience.
\end{remark}
\begin{remark}
Under DCDFM, the intuition of considering connectivity matrix $P$ comes from the fact we need a block matrix to generate $\mathbb{E}[A]$, similar as the SBM and DCSBM models. In detail, under DCDFM, if we do not consider connectivity matrix $P$ (i.e., if $P$ is an identity matrix), since $\mathbb{E}[A(i,j)]=\Omega(i,j)=\theta(i)\theta(j)P(\ell(i),\ell(j))$, we have $\mathbb{E}[A(i,j)]=0$ if nodes $i$ and $j$ are in different communities, and this limits the popularity of the applicability of a model. Therefore, we need to consider a connectivity matrix in our model. Note that $P$ is not a matrix with probabilities unless $\mathcal{F}$ is Bernoulli or Poisson or Binomial distributions. See, when $\mathcal{F}$ is Normal distribution, we can let $P$ has negative values such that $\mathbb{E}[A(i,j)]=\theta(i)\theta(j)P(\ell(i),\ell(j))$, i.e., $A$ can have negative values. For example, \cite{airoldi2013multi} generates its multi-way blockmodels by letting their adjacency matrix generated from a Normal distribution with a block matrix which can have negative entries.
\end{remark}
\begin{remark}
Relationship between DCDFM and DFM is similar as that between DCSBM and SBM. When $\Theta=\sqrt{\rho}I_{n}$ for $\rho>0$, we have $\mathbb{E}[A(i,j)]=\rho P(\ell(i),\ell(j))$, i.e., $A(i,j)$'s expectation only depends on $\rho P(\ell(i),\ell(j))$; instead, if $\Theta\neq \sqrt{\rho}I_{m}$, we have $\mathbb{E}[A(i,j)]=\theta(i)\theta(j)P(\ell(i),\ell(j))$, i.e., $A(i,j)$'s expectation can be modeled by not only the community information but also the individual characters of nodes $i$ and $j$. Thus, we see that by considering $\theta$, DCDFM is more applicable than DFM since DCDFM considers node individuality.
\end{remark}
\begin{remark}
When $\mathcal{F}$ is Bernoulli or Poisson or Binomial distributions, all entries of $A$ take values from $\{0,1,2,\ldots,m\}$ for some positive integer $m$, there may exist a subset of nodes never connect in $A$ since many entries of $A$ are 0 while other entries are positive integers, i.e., $A$ may be dis-connected and this happens when the network is sparse. However, when  $\mathcal{F}$ is a distribution of continuous random variables (for example, when $\mathcal{F}$ is Normal distribution), all node pairs are connected and $A$ is a connected matrix naturally.
\end{remark}
\begin{remark}
Our model DCDFM is applicable for network in which nodes may be self-connected, and this is also verified by the proof of Lemma 2 which has no restriction on diagonal entries of A.
\end{remark}
Next proposition guarantees the identifiability of DCDFM, and such identifiability is similar as that of DCSBM.
\begin{proposition}\label{idDCDFM}
(Identifiability of DCDFM).	DCDFM is identifiable for membership matrix: For eligible  $(P,Z, \Theta)$ and $(\tilde{P}, \tilde{Z},\tilde{\Theta})$, if $\Theta ZPZ'\Theta=\tilde{\Theta}\tilde{Z}\tilde{P}\tilde{Z}'\tilde{\Theta}$, then $Z=\tilde{Z}$.
\end{proposition}
\begin{proof}
Set $\tilde{\ell}(i)=k$ if $\tilde{Z}(i,k)=1$ for $i\in[n],k\in[K]$. By Lemma \ref{GenUrUc}, under DCDFM, $U_{*}=ZB$ gives $U_{*}(i,:)=Z(i,:)B=B(\ell(i),:)=\tilde{Z}(i,:)B=B(\tilde{\ell}(i),:)$, which gives $Z=\tilde{Z}$.
\end{proof}
Compared with the DCSBM of \cite{DCSBM}, our model DCDFM has no distribution constraint on all entries of $A$ and allows $P$ to have negative entries while DCSBM requires that $A$ follows Bernoulli or Poisson distribution and all  entries of $P$ are nonnegative. Such differences make that our DCDFM can model both un-weighted networks and weighted networks while DCSBM only models un-weighted networks. Sure, DCDFM is a direct extension of DFM by considering node heterogeneity. Meanwhile, though the WSBM introduced in \cite{palowitch2018significance} also considers node heterogeneity, it requires all entries of $P$ to be nonnegative, and this limits the generality of WSBM.
\section*{Algorithm: nDFA}
We will now introduce our inference algorithm aiming at estimating $\ell$ given $A$ and $K$ under DCDFM. Since  $\mathrm{rank}(P)=K, \mathrm{rank}(Z)=K$ and $\mathrm{rank}(\Omega)=K$, $\Omega$ has $K$ nonzero eigenvalues by basic algebra. Let $\Omega=U\Lambda U'$ be the compact eigenvalue decomposition of $\Omega$ with $U\in\mathbb{R}^{n\times K}, \Lambda\in\mathbb{R}^{K\times K}$, and $U'U=I_{K}$ where $I_{K}$ is a $K\times K$ identity matrix. Let $U_{*}\in \mathbb{R}^{n\times K}$ be the row-normalized version of $U$ such that $U_{*}(i,:)=\frac{U(i,:)}{\|U(i,:)\|_{F}}$ for $i\in[n]$.  Let $\mathcal{I}$ be the indices of nodes corresponding to $K$ communities, one from each community. The following lemma provides the intuition about designing our algorithm to fit the proposed model.
\begin{lemma}\label{GenUrUc}
Under $DCDFM_{n}(K,P,Z,\Theta)$, we have $U_{*}=ZB$, where $B=U_{*}(\mathcal{I},:)$.
\end{lemma}
\begin{proof}
The facts $\Omega=\Theta ZPZ'\Theta=U\Lambda U'$ and $U'U=I_{K}$ give $U=\Theta ZPZ'\Theta U\Lambda^{-1}=\Theta Z\tilde{B}$ where we set $\tilde{B}=PZ'\Theta U\Lambda^{-1}$ for convenience. For $i\in[n]$, $U=\Theta Z\tilde{B}$ gives $U(i,:)=\theta(i)Z(i,:)\tilde{B}=\theta(i)\tilde{B}(\ell(i),:)$, then we have $U_{*}(i,:)=\frac{U(i,:)}{\|U(i,:)\|_{F}}=\frac{\tilde{B}(\ell(i),:)}{\|\tilde{B}(\ell(i),:)\|_{F}}$. When $\ell(i)=\ell(\bar{i})$, we have $U_{*}(i,:)=U_{*}(\bar{i},:)$, and this gives $U_{*}=ZB$, where $B=U_{*}(\mathcal{I},:)$.
\begin{remark}\label{whythetapositive}
In the proof of Lemma \ref{GenUrUc}, we see that to make $U_{*}(i,:)=\frac{U(i,:)}{\|U(i,:)\|_{F}}=\frac{\tilde{B}(\ell(i),:)}{\|\tilde{B}(\ell(i),:)\|_{F}}$ hold such that $U_{*}(i,:)=U_{*}(\bar{i},:)$ holds if $\ell(i)=\ell(\bar{i})$, Eq (\ref{DefinTheta}) should hold. Otherwise, if some entries of $\theta$ are negative while others are positive, we have $U_{*}(i,:)=\frac{U(i,:)}{\|U(i,:)\|_{F}}=\frac{\theta(i)}{|\theta(i)|}\frac{\tilde{B}(\ell(i),:)}{\|\tilde{B}(\ell(i),:)\|_{F}}$, which gives that $U_{*}(i,:)=U_{*}(\bar{i},:)$ may not hold when $\ell(i)=\ell(\bar{i})$ unless we assume that $\frac{\theta(i)}{|\theta(i)|}=\frac{\theta(\bar{i})}{|\theta(\bar{i})|}$ when $\ell(i)=\ell(\bar{i})$.
\end{remark}
\end{proof}
Lemma \ref{GenUrUc} says that rows of $U_{*}$ corresponding to nodes of the same clusters are equal. This suggests that applying k-means algorithm on all rows of $U_{*}$ assuming there are $K$ communities exactly returns nodes memberships  up to a permutation of nodes labels since $U_{*}$ has $K$ different rows and $U_{*}(i,:)=U_{*}(\bar{i},:)$ if nodes $i$ and $\bar{i}$ belong to the same community for $i,\bar{i}\in[n]$.

The above analysis is under the oracle case when $\Omega$ is given under DCDFM, now we turn to the real case where we only have $A$ obtained from the weighted network $\mathcal{N}$ and the known number of communities $K$. Since labels vector $\ell$ is unknown for the real case, our goal is to use $(A,K)$ to predict it. Let $\tilde{A}=\hat{U}\hat{\Lambda}\hat{U}'$ be the leading $K$ eigen-decomposition of $A$ such that $\hat{U}\in \mathbb{R}^{n\times K}, \hat{\Lambda}\in \mathbb{R}^{K\times K}, \hat{U}'\hat{U}=I_{K}$, and $\hat{\Lambda}$ contains the leading $K$ eigenvalues of $A$. Let $\hat{U}_{*}\in \mathbb{R}^{n\times K}$ be the row-normalized version of $\hat{U}$ such that $\hat{U}_{*}(i,:)=\frac{\hat{U}(i,:)}{\|\hat{U}(i,:)\|_{F}}$ for $i\in[n]$. The detail of our normalized Distribution-Free Algorithm (nDFA for short) is described in Algorithm \ref{alg:nDFA}, and it can be programmed by only a few lines of Matlab codes.
\begin{algorithm}
	\caption{normalized Distribution-Free Algorithm}
	\label{alg:nDFA}
	\begin{algorithmic}[1]
		\Require Adjacency matrix $A\in \mathbb{R}^{n\times n}$, and number of communities $K$.
		\Ensure The estimated $n\times 1$ labels vector $\hat{\ell}$.
		\State Let $\tilde{A}=\hat{U}\hat{\Lambda}\hat{U}'$ be the leading $K$ eigen-decomposition of $A$ such that $\hat{\Lambda}$ contains the leading $K$ eigenvalues of $A$ and $\hat{U}'\hat{U}=I_{K}$.
		\State Let $\hat{U}_{*}$ be the row-normalization of $\hat{U}$.
		\State Apply k-means on all rows of $\hat{U}_{*}$ with $K$ clusters to obtain $\hat{\ell}$.
	\end{algorithmic}
\end{algorithm}
We name our algorithm as nDFA to stress the normalization procedure aiming at cancelling the effect of node heterogeneity and the distribution-free property aiming at modeling weighted networks. Using the idea of normalizing each row of $\hat{U}$ to remove the effect of node heterogeneity can also be found in \cite{RSC,lei2015consistency}. Using the idea of entry-wise ratios between the leading eigenvector and other leading eigenvectors of $A$ proposed in \cite{SCORE} is also possible to remove the effect of $\Theta$, and we leave studies of it under DCDFM for our future work.

Here, we provide the complexity analysis of nDFA. For nDFA's computational complexity, the most expensive step is the eigenvalue decomposition which requires $O(n^{3})$ times \cite{tsironis2013accurate}. The row-normalization step costs $O(n^{2})$, and the k-means step costs $O(nlK^{2})$, where $l$ is the number of k-means iterations, and we set $l=100$ for our nDFA in this article. So the overall computational complexity of nDFA is $O(n^{3})$. Though it is time demanding when $n$ becomes huge, many wonderful works focus on spectral clustering for un-weighted network community detection, see \cite{rohe2011spectral,RSC,joseph2016impact,lei2015consistency,SCORE,rohe2016co,MixedSCORE,SPACL,OCCAM,MaoSVM,SLIM,zhou2019analysis,DSCORE}.
Meanwhile, though using the random-projection and random-sampling ideas developed in \cite{zhang2022randomized} to accelerate nDFA is possible, it is out of the scope of this article, and we leave it for our future work.
\section*{Consistency of nDFA}
To build theoretical guarantee on nDFA's consistency under DCDFM, we need below assumption.
\begin{assumption}\label{assumeVar}
Assume
\begin{itemize}
  \item $\tau=\mathrm{max}_{i,j\in[n]}|A(i,j)-\Omega(i,j)|$ is finite.
  \item $\gamma=\mathrm{max}_{i,j\in[n]}\frac{\mathrm{Var}(A(i,j))}{\theta(i)\theta(j)}$ is finite.
\end{itemize}
\end{assumption}
The above assumption is mild since it only requires that all elements of $A$ and $\Omega$, and variances of $A$'s entries are finite. We'd emphasize that Assumption \ref{assumeVar} has no prior knowledge on any specific distribution of $A(i,j)$ under DCDFM for all nodes, thus it dose not violate the distribution-free property of the proposed model. To build theoretical guarantee on consistent estimation, we need the following assumption.
\begin{assumption}\label{assumesparsity}
Assume $\gamma\theta_{\mathrm{max}}\|\theta\|_{1}\geq \tau^{2}\mathrm{log}(n)$.
\end{assumption}
On the one hand, when all elements of $A$ are nonnegative, Assumption \ref{assumesparsity} guarantees a lower bound requirement on network sparsity.  To have a better understanding on network sparsity, consider the case that $\mathcal{F}$ is a distribution such that all entries of $A$ are nonnegative. We have $\sum_{j=1}^{n}A(i,j)$ is the degree of node $i$ and $\sum_{j=1}^{n}\Omega(i,j)=\theta(i)\sum_{j=1}^{n}\theta(j)P(\ell(i),\ell(j))$ is the expectation degree of node $i$. Especially, when $\Theta=\sqrt{\rho}I_{n}$ and $\mathcal{F}$ is Bernoulli or Poisson or Binomial distribution, we have $\Omega=\rho ZPZ'$, which gives $\mathbb{P}(A(i,j)=m)=\rho P(\ell(i),\ell(j))$ for some $m>0$, we see that $\rho$ controls the sparsity of such weighted network or un-weighted network. Meanwhile, the sparsity assumption is common when proving estimation consistency for spectral clustering method, for example, consistency works for un-weighted network community detection like \cite{lei2015consistency,SCORE}. Especially, when $\mathcal{F}$ is Bernoulli distribution and $\Theta=\sqrt{\rho}I_{n}$ such that DCDFM reduces to SBM, $\gamma$ and $\tau$ have a upper bound 1, and Assumption \ref{assumesparsity} turns to require that $\rho n\geq \mathrm{log}(n)$, which is consistent with the sparsity requirement under SBM in \cite{lei2015consistency}, and this guarantees that our requirement on network sparsity matches with classical result when DCDFM degenerates to SBM. On the other hand, for the case that $\mathcal{F}$ allows $A$ to have negative entries, $\theta$ is not related with network sparsity but only heterogeneity parameter because it is meaningless to define sparsity in an adjacency matrix with negative elements. For this case, Assumption \ref{assumesparsity} merely controls $\theta$ for our theoretical framework.

Though $\gamma$ is assumed to be a finite number, we also consider it in our Assumption \ref{assumesparsity} due to the fact that $\gamma$ is directly related with the variance term of $A$'s elements, i.e., $\gamma$ has a close relationship with the distribution $\mathcal{F}$ though $\mathcal{F}$ can be arbitrary distribution. After obtaining our main results for nDFA, we will apply some examples to show that $\tau$ and $\gamma$ are always finite or at least can be set as finite numbers to make Assumption \ref{assumesparsity} hold under different choices of  $\mathcal{F}$. Meanwhile, we make Assumptions \ref{assumeVar} and \ref{assumesparsity} mainly for the convenience of theoretical analysis on nDFA's consistent estimation, and this two assumptions are irrelevant to the identifiability of our model DCDFM. Based on the above two assumptions, the following lemma bounds $\|A-\Omega\|$ with an application of Bernstein inequality \cite{tropp2012user}.
\begin{lemma}\label{BoundAOmega}
Under $DCDFM_{n}(K,P,Z,\Theta)$, when Assumptions \ref{assumeVar} and \ref{assumesparsity} hold, with probability at least $1-o(n^{-3})$, we have
\begin{align*}
\|A-\Omega\|=O(\sqrt{\gamma\theta_{\mathrm{max}}\|\theta\|_{1}\mathrm{log}(n)}).
\end{align*}
\end{lemma}
\begin{proof}
Set $H=A-\Omega$, we have $H=\sum_{i=1}^{n}\sum_{j=1}^{n}H(i,j)e_{i}e'_{j}$. Set $H^{(i,j)}=H(i,j)e_{i}e'_{j}$. Since $\mathbb{E}[H(i,j)]=\mathbb{E}[A(i,j)-\Omega(i,j)]=0$, we have $\mathbb{E}[H^{(i,j)}]=0$ and
\begin{align*}
\|H^{(i,j)}\|&=\|(A(i,j)-\Omega(i,j))e_{i}e'_{j}\|=|A(i,j)-\Omega(i,j)|\|e_{i}e'_{j}\|=|A(i,j)-\Omega(i,j)|\leq\tau,
\end{align*}
where the last inequality holds by Assumption \ref{assumeVar}. Set $R=\tau$, and $\sigma^{2}=\|\sum_{i=1}^{n}\sum_{j=1}^{n}\mathbb{E}[H^{(i,j)}(H^{(i,j)})']\|$. Since $\mathbb{E}[H^{2}(i,j)]=\mathbb{E}[(A(i,j)-\Omega(i,j))^{2}]=\mathrm{Var}(A(i,j))\leq \gamma\theta(i)\theta(j)$ by Assumption \ref{assumeVar}, we have
\begin{align*}
\|\sum_{i=1}^{n}\sum_{j=1}^{n}\mathbb{E}(H^{(i,j)}(H^{(i,j)})')\|&=\|\sum_{i=1}^{n}\sum_{j=1}^{n}\mathbb{E}(H^{2}(i,j))e_{i}e'_{i}\|\leq \gamma\theta_{\mathrm{max}}\|\theta\|_{1},
\end{align*}
which gives $\sigma^{2}\leq \gamma\theta_{\mathrm{max}}\|\theta\|_{1}$. Set $t=\frac{\alpha+1+\sqrt{\alpha^{2}+20\alpha+19}}{3}\sqrt{\gamma\theta_{\mathrm{max}}\|\theta\|_{1}\mathrm{log}(n)}$. Theorem 1.4 (the Matrix Bernstein) of \cite{tropp2012user} gives
\begin{align*}
\mathbb{P}(\|H\|\geq t)&\leq n\mathrm{exp}(-\frac{t^{2}/2}{\sigma^{2}+\frac{Rt}{3}})\leq n\mathrm{exp}(-\frac{t^{2}/2}{(\gamma\theta_{\mathrm{max}}\|\theta\|_{1})+\frac{Rt}{3}})=n\mathrm{exp}(-\frac{(\alpha+1)\mathrm{log}(n)}{\frac{18}{(\sqrt{\alpha+19}+\sqrt{\alpha+1})^{2}}+\frac{2\sqrt{\alpha+1}}{\sqrt{\alpha+19}+\sqrt{\alpha+1}}\sqrt{\frac{R^{2}\mathrm{log}(n)}{\gamma\theta_{\mathrm{max}}\|\theta\|_{1}}}})\\
&\leq n\mathrm{exp}(-(\alpha+1)\mathrm{log}(n))=\frac{1}{n^{\alpha}},
\end{align*}
where the last inequality comes from Assumption \ref{assumesparsity}. Set $\alpha=3$, the claim follows.
\end{proof}
Bound obtained in Lemma \ref{BoundAOmega} is directly related with our main result for nDFA. To measure nDFA's performance theoretically, we apply the clustering error of \cite{joseph2016impact} for its theoretical convenience. Set $\hat{\mathcal{C}}_{k}=\{i: \hat{\ell}(i)=k\}$ for $k\in[K]$. Define the clustering error as
\begin{align*}
\hat{f}=\mathrm{min}_{\pi\in S_{K}}\mathrm{max}_{k\in[K]}\frac{|\mathcal{C}_{k}\cap \hat{\mathcal{C}}^{c}_{\pi(k)}|+|\mathcal{C}^{c}_{k}\cap \hat{\mathcal{C}}_{\pi(k)}|}{n_{k}},
\end{align*}
where $S_{K}$ is the set of all permutations of $\{1,2,\ldots,K\}$. Actually, using clustering errors in \cite{RSC,lei2015consistency, SCORE} to measure nDFA's performance also works, and we use $\hat{f}$ in this paper mainly for its convenience in proofs. The  following theorem is the main result for nDFA, and it shows that nDFA enjoys asymptotically consistent estimation under the proposed model.
\begin{theorem}\label{MainnDFA}
Under $DCDFM_{n}(K,P,Z,\Theta)$, let $\hat{\ell}$ be obtained from nDFA, when Assumptions \ref{assumeVar} and \ref{assumesparsity} hold, with probability at least $1-o(n^{-3})$, we have
\begin{align*}
\hat{f}=O(\frac{\gamma\theta^{3}_{\mathrm{max}}K^{2}n_{\mathrm{max}}\|\theta\|_{1}\mathrm{log}(n)}{\lambda^{2}_{K}(P)\theta^{6}_{\mathrm{min}}n^{3}_{\mathrm{min}}}).
\end{align*}
\end{theorem}
\begin{proof}
The following lemma provides a general lower bound of $\sigma_{K}(\Omega)$ under DCDFM, where this lower bound is directly related with model parameters.
\begin{lemma}\label{svdK}
Under $DCDFM_{n}(K,P,Z,\Theta)$, we have $|\lambda_{K}(\Omega)|\geq\theta^{2}_{\mathrm{min}}|\lambda_{K}(P)|n_{\mathrm{min}}$.
\end{lemma}
\begin{proof}
\begin{align*}
\lambda_{K}(\Omega\Omega')&=\lambda_{K}(\Theta ZPZ'\Theta^{2}ZP'Z'\Theta)=\lambda_{K}(\Theta^{2}ZPZ'\Theta^{2}ZP'Z')\geq\theta^{2}_{\mathrm{min}}\lambda_{K}(Z'ZPZ'\Theta^{2}ZP')\geq \theta^{2}_{\mathrm{min}}\lambda_{K}(Z'Z)\lambda_{K}(PZ'\Theta^{2}ZP')\\
&=\theta^{2}_{\mathrm{min}}\lambda_{K}(Z'Z)\lambda_{K}(Z'\Theta^{2}ZP'P)\geq\theta^{2}_{\mathrm{min}}\lambda_{K}(Z'Z)\lambda_{K}(Z'\Theta^{2}Z)\lambda_{K}(PP')\geq\theta^{4}_{\mathrm{min}}\lambda^{2}_{K}(Z'Z)\lambda_{K}(PP')=\theta^{4}_{\mathrm{min}}n^{2}_{\mathrm{min}}|\lambda_{K}(P)|^{2},
\end{align*}
where we have used the fact that $\lambda_{K}(Z'Z)=n_{\mathrm{min}}$ in the last equality.
\end{proof}
By Lemma 5.1 of \cite{lei2015consistency}, there is a $K\times K$ orthogonal matrix $\mathcal{O}$ such that
\begin{align*}
\|\hat{U}\mathcal{O}-U\|_{F}\leq \frac{2\sqrt{2K}\|A-\Omega\|}{|\lambda_{K}(\Omega)|}\leq\frac{2\sqrt{2K}\|A-\Omega\|}{\theta^{2}_{\mathrm{min}}|\lambda_{K}(P)|n_{\mathrm{min}}},
\end{align*}
where we have applied Lemma \ref{svdK} in the last inequality.

For $i\in[n]$, by basic algebra, we have
$\|\hat{U}_{*}(i,:)\mathcal{O}-U_{*}(i,:)\|_{F}\leq \frac{2\|\hat{U}(i,:)\mathcal{O}-U(i,:)\|_{F}}{\|U(i,:)\|_{F}}$.
Set $m_{U}=\mathrm{min}_{i\in[n]}\|U(i,:)\|_{F}$, we have
\begin{align*}
\|\hat{U}_{*}\mathcal{O}-U_{*}\|_{F}&=\sqrt{\sum_{i=1}^{n}\|\hat{U}_{*}(i,:)\mathcal{O}-U_{*}(i,:)\|^{2}_{F}}\leq \frac{2\|\hat{U}\mathcal{O}-U\|_{F}}{m_{U}}.
\end{align*}
According to the proof of Lemma 3.5 of \cite{qing2021consistency} where this lemma is distribution-free, we have $\frac{1}{m_{U}}\leq \frac{\theta_{\mathrm{max}}\sqrt{n_{\mathrm{max}}}}{\theta_{\mathrm{min}}}$, which gives
\begin{align*}
\|\hat{U}_{*}\mathcal{O}-U_{*}\|_{F}&\leq \frac{2\|\hat{U}\mathcal{O}-U\|_{F}}{m_{U}}\leq
\frac{4\theta_{\mathrm{max}}\sqrt{2Kn_{\mathrm{max}}}\|A-\Omega\|}{|\lambda_{K}(P)|\theta^{3}_{\mathrm{min}}n_{\mathrm{min}}}.
\end{align*}

By Lemma 2 in \cite{joseph2016impact}, for each $1\leq k\neq l\leq K$, if one has
\begin{align}\label{holdGIDCSBM}
\frac{\sqrt{K}}{\varsigma}\|U_{*}-\hat{U}_{*}\mathcal{O}\|_{F}(\frac{1}{\sqrt{n_{k}}}+\frac{1}{\sqrt{n_{l}}})\leq \|B(k,:)-B(l,:)\|_{F},
\end{align}
then $\hat{f}=O(\varsigma^{2})$. Since  $\|B(k,:)-B(l,:)\|_{F}=\sqrt{2}$ for any $k\neq l$ by Lemma 3.1 of \cite{qing2021consistency}, setting $\varsigma=\frac{2}{\sqrt{2}}\sqrt{\frac{K}{n_{\mathrm{min}}}}\|U_{*}-\hat{U}_{*}\mathcal{O}\|_{F}$ makes Eq (\ref{holdGIDCSBM}) hold for all $1\leq k\neq l\leq K$, and then we have $\hat{f}=O(\varsigma^{2})=O(\frac{K\|U_{*}-\hat{U}_{*}\mathcal{O}\|^{2}_{F}}{n_{\mathrm{min}}})$. Thus,$\hat{f}=O(\frac{\theta^{2}_{\mathrm{max}}K^{2}n_{\mathrm{max}}\|A-\Omega\|^{2}}{\lambda^{2}_{K}(P)\theta^{6}_{\mathrm{min}}n^{3}_{\mathrm{min}}})$.
Finally, this theorem follows by Lemma \ref{BoundAOmega}.
\end{proof}
From Theorem \ref{MainnDFA}, we see that decreasing $\theta_{\mathrm{min}}$ increases the upper bound of error rate, and this can be understood naturally since a smaller $\theta_{\mathrm{min}}$ gives a higher probability to generate an isolated node having no connections with other nodes, and thus a harder case for community detection, where such fact is also found in \cite{SCORE} under DCSBM. It is also harder to detect nodes labels for a network generated under a smaller $|\lambda_{K}(P)|$ and $n_{\mathrm{min}}$, and such facts are also found in \cite{lei2015consistency} under DCSBM. Add some conditions on model parameters, we have below corollary by basic algebra.
\begin{corollary}\label{cor}
Under $DCDFM_{n}(K,P,Z,\Theta)$, and conditions in Theorem \ref{MainnDFA} hold, with probability at least $1-o(n^{-3})$,
\begin{itemize}
  \item when $K=O(1), \frac{n_{\mathrm{max}}}{n_{\mathrm{min}}}=O(1)$, we have $
\hat{f}=O(\frac{\gamma\theta^{3}_{\mathrm{max}}\|\theta\|_{1}\mathrm{log}(n)}{\lambda^{2}_{K}(P)\theta^{6}_{\mathrm{min}}n^{2}})$.
  \item when $\theta_{\mathrm{max}}=O(\sqrt{\rho}), \theta_{\mathrm{min}}=O(\sqrt{\rho})$ for $\rho>0$, we have $\hat{f}=O(\frac{\gamma K^{2}n_{\mathrm{max}}n\mathrm{log}(n)}{\lambda^{2}_{K}(P)\rho n^{3}_{\mathrm{min}}})$.
  \item  when $K=O(1), \frac{n_{\mathrm{max}}}{n_{\mathrm{min}}}=O(1), \theta_{\mathrm{max}}=O(\sqrt{\rho})$ and $\theta_{\mathrm{min}}=O(\sqrt{\rho})$ for $\rho>0$, we have $\hat{f}=O(\frac{\gamma\mathrm{log}(n)}{\lambda^{2}_{K}(P)\rho n})$.
\end{itemize}
\end{corollary}
When $\theta(i)=\sqrt{\rho}$ for $i\in[n]$ such that DCDFM reduces to DFM, theoretical results for nDFA under DCDFM are consistent with those under DFM proposed in Theorem 1 of \cite{qing2021DFM}. For the third bullet of Corollary \ref{cor}, we see that $|\lambda_{K}(P)|$ should shrink slower than $\sqrt{\frac{\gamma\mathrm{log}(n)}{\rho n}}$ for consistent estimation, and it should shrink slower than $\sqrt{\frac{\mathrm{log}(n)}{n}}$ when $\rho$ is a constant and $\gamma$ is finite. When $\lambda_{K}(P)$ and $\gamma$ are fixed, we see that $\rho$ should shrink slower than $\frac{\mathrm{log}(n)}{n}$, and this is consistent with assumption \ref{assumesparsity}. Generally speaking, the finiteness of $\gamma$ is significant for the fact that we can ignore the effect of $\gamma$ in our theoretical bounds as long as $\gamma$ is finite. Next, we use some examples under different distributions to show that $\tau$ and $\gamma$ are finite or we can always set them as finite.

Follow similar analysis as \cite{qing2021DFM}, we let $\mathcal{F}$ be some specific distributions as examples to show the generality of DCDFM as well as nDFA's consistent estimation under DCDFM. For $i,j\in[n]$, we mainly bound $\gamma$ to show that $\gamma$ is finite (i.e., the 2nd bullet in assumption \ref{assumeVar} holds under different distributions) and then obtain error rates of nDFA by considering below distributions under DCDFM, where details on probability mass function or probability density function on these distributions can be found in \url{http://www.stat.rice.edu/~dobelman/courses/texts/distributions.c&b.pdf}.
\begin{Ex}\label{Normal}
when $\mathcal{F}$ is Normal distribution such that  $A(i,j)\sim \mathrm{Normal}(\Omega(i,j),\sigma^{2}_{A})$ and all entries of $A$ are finite real numbers. Since mean of Normal distribution can be negative, DCDFM allows $P$ to have negative entries as long as $P$ is full rank. Sure, in this case, $\gamma=\mathrm{max}_{i,j\in[n]}\frac{\sigma^{2}_{A}}{\theta(i)\theta(j)}\leq \frac{\sigma^{2}_{A}}{\theta^{2}_{\mathrm{min}}}$ is finite, and assumption \ref{assumesparsity} requires $\frac{\sigma^{2}_{A}}{\theta^{2}_{\mathrm{min}}}\frac{\theta_{\mathrm{max}}\|\theta\|_{1}}{\mathrm{log}(n)}\rightarrow\infty$ as $n\rightarrow\infty$. Set $\gamma=O(\frac{\sigma^{2}_{A}}{\theta^{2}_{\mathrm{min}}})$ in Theorem \ref{MainnDFA}, nDFA's error bound is
\begin{align*}
\hat{f}=O(\frac{\sigma^{2}_{A}}{\theta^{2}_{\mathrm{min}}}\frac{\theta^{3}_{\mathrm{max}}K^{2}n_{\mathrm{max}}\|\theta\|_{1}\mathrm{log}(n)}{\lambda^{2}_{K}(P)\theta^{6}_{\mathrm{min}}n^{3}_{\mathrm{min}}}).
\end{align*}
From this bound, we see that increases $\sigma^{2}_{A}$ increases error rate, and a smaller $\sigma^{2}_{A}$ is preferred which is also verified by Experiment 1[b] in Section \ref{ExperimentalStud}. For convenience, setting $\sigma^{2}_{A}\leq C\theta^{2}_{\mathrm{min}}$ for some $C>0$ makes assumption \ref{assumesparsity} equal to require $\frac{\theta_{\mathrm{max}}\|\theta\|_{1}}{\mathrm{log}(n)}\rightarrow\infty$ as $n\rightarrow\infty$ since $\tau$ is finite and $\hat{f}=O(\frac{\theta^{3}_{\mathrm{max}}K^{2}n_{\mathrm{max}}\|\theta\|_{1}\mathrm{log}(n)}{\lambda^{2}_{K}(P)\theta^{6}_{\mathrm{min}}n^{3}_{\mathrm{min}}})$.
\end{Ex}
\begin{Ex}\label{Binomial}
When $\mathcal{F}$ is Binomial distribution such that  $A(i,j)\sim \mathrm{Binomial}(m,\frac{\Omega(i,j)}{m})$ for some positive integer $m$ and all entries of $A$ are integers in $\{0,1,2,\ldots,m\}$. Sure, $\tau\leq m$ here. For this case, since $\frac{\Omega(i,j)}{m}$ is probability, all elements of $P$ should be nonnegative. Sure, we have $\mathbb{E}[A(i,j)]=\Omega(i,j)$ and $\mathrm{Var}(A(i,j))=\Omega(i,j)(1-\frac{\Omega(i,j)}{m})\leq \Omega(i,j)\leq\theta(i)\theta(j)$ by the property of Binomial distribution. Thus, $\gamma=1$, and error rate in this case can be obtained immediately.
\end{Ex}
\begin{Ex}\label{Bernoulli}
When $\mathcal{F}$ is Bernoulli distribution, we have $A(i,j)\sim\mathrm{Bernoulli}(\Omega(i,j))$, all entries of $P$ are nonnegative, and DCDFM reduces to DCSBM considered in literature \cite{SCORE, lei2015consistency}. For this case, all entries of $A$ are either 0 or 1, i.e., un-weighted network and $\tau\leq 1$. Since $A(i,j)\sim\mathrm{Bernoulli}(\Omega(i,j))$ and $\Omega(i,j)$ is a probability in $[0,1]$, we have $\mathrm{Var}(A(i,j))=\Omega(i,j)(1-\Omega(i,j))\leq \Omega(i,j)=\theta(i)\theta(j)P(\ell(i),\ell(j))\leq \theta(i)\theta(j)$, suggesting that $\gamma=1$ is finite and Assumption \ref{assumeVar} holds. Setting $\gamma=1$ in Theorem \ref{MainnDFA} obtains the theoretical upper bound of nDFA's error rate under DCDFM  immediately.
\end{Ex}
\begin{Ex}\label{Poisson}
when $\mathcal{F}$ is Poisson distribution such that  $A(i,j)\sim \mathrm{Poisson}(\Omega(i,j))$ as in \cite{DCSBM} and all entries of $A$ are nonnegative integers. For Poisson distribution, all entries of $P$ should be nonnegative and $\tau$ is finite as long as $A$'s elements are generated from Poisson distribution. Meanwhile, $\mathbb{E}[A(i,j)]=\Omega(i,j)$ and $\mathrm{Var}(A(i,j))=\Omega(i,j)\leq \theta(i)\theta(j)$ holds by Poisson distribution's expectation and variance. Thus, $\gamma=1$ is finite and we can obtain bound of nDFA's error rate from Theorem \ref{MainnDFA}. For this example, under conditions in the 3rd bullet of Corollary \ref{cor}, our theoretical result matches that of Theorem 4.2 \cite{lei2015consistency} up to logarithmic factors, and this guarantees the optimality of our theoretical studies.
\end{Ex}
\begin{Ex}\label{Logistic}
When $\mathcal{F}$ is Logistic distribution such that $A(i,j)\sim\mathrm{Logistic}(\Omega(i,j),\beta)$ for $\beta>0$ and all entries of $A$ are real values.  For this case, $P$'s entries are real values, $\tau$ is finite when $A$'s entries come from Logistic distribution, $\mathbb{E}[A(i,j)]=\Omega(i,j)$ satisfying Eq (\ref{DefinMM}), and $\mathrm{Var}(A(i,j))=\frac{\pi^{2}\beta^{2}}{3}$, i.e., $\gamma\leq\frac{\pi^{2}\beta^{2}}{3\theta^{2}_{\mathrm{min}}}$ and $\gamma$ is finite.
\end{Ex}
\begin{Ex}\label{Signed}
DCDFM can also generate signed network by setting $\mathbb{P}(A(i,j)=1)=\frac{1+\Omega(i,j)}{2}$ and $\mathbb{P}(A(i,j)=-1)=\frac{1-\Omega(i,j)}{2}$ such that all non-diagonal elements of $A$ are either $1$ or $-1$. For this case, all entries of $P$ are real values and $\Omega(i,j)$ should be set such that $-1\leq \Omega(i,j)\leq1$. Sure, $\mathbb{E}[A(i,j)]=\Omega(i,j)$ satisfies Eq (\ref{DefinMM}), and $\mathrm{Var}(A(i,j))=1-\Omega^{2}(i,j)\leq1$, i.e., $\gamma\leq \frac{1}{\theta^{2}_{\mathrm{min}}}$ is finite.
\end{Ex}
Since our model DCDFM has no limitation on the choice of distribution $\mathcal{F}$ as long as Eq (\ref{DefinMM}) holds, setting $\mathcal{F}$ as any other distribution (see, Double exponential, Exponential, Gamma and Uniform distributions in \url{http://www.stat.rice.edu/~dobelman/courses/texts/distributions.c&b.pdf}) obeying Eq (\ref{DefinMM}) is also possible and this guarantees the generality of our model as well as our theoretical results.
\section*{Experimental Results}\label{ExperimentalStud}
Both simulated and empirical data are presented to compare nDFA with existing algorithm DFA developed in \cite{qing2021DFM} for weighted networks, where DFA applies k-means on all rows of $\hat{U}$ with $K$ clusters to estimate nodes labels. Meanwhile, codes for all experimental results in this paper are executed by MATLAB R2021b. Though our model DCDFM and our algorithm nDFA are also applicable for network with self-connected nodes, unless specified, we do not consider loops in this part. Before presenting experimental results, we introduce general modularity for weighted network community detection in next subsection, where the general modularity can be seen as a measure of performance for any algorithm designed for weighted network and we will also test the effectiveness of the general modularity in both simulated and empirical networks.
\subsection*{General modularity for weighted networks}
Unlike un-weighted network, the node degree in weighted network is slightly different, especially when $A$ has negative elements. For un-weighted network in which all entries of $A$ take values either 0 or 1, and weighted network in which all entries of $A$ are nonnegative, degree for node $i$ is always defined as $\sum_{j=1}^{n}A(i,j)$. However, for weighted network in which $A$ has negative entries, $\sum_{j=1}^{n}A(i,j)$ does not measure the degree of node $i$.  Instead, to measure node degree for all kinds of weighted networks, we define degree of node $i$ as below: let $A^{+}, A^{-}\in\mathbb{R}^{n\times n}$ such that $A^{+}(i,j)=\mathrm{max}(A(i,j),0)$ and $A^{-}(i,j)=\mathrm{max}(-A(i,j),0)$ for all $i,j$. Then we have $A=A^{+}-A^{-}$. Let $d^{+}(i)=\sum_{j=1}^{n}A^{+}(i,j)$ and $d^{-}(i)=\sum_{j=1}^{n}A^{-}(i,j)$ be the positive and negative degrees of node $i$. Meanwhile, since $\mathbb{E}[\sum_{j=1}^{n}|A(i,j)|]=\sum_{j=1}^{n}\mathbb{E}[|A(i,j)|]=\theta(i)\sum_{j=1}^{n}\theta(j)|P(\ell(i),\ell(j))|$, we see that $\theta(i)$ is a measure of the ``degree" of node $i$, especially when all entries of $A$ are nonnegative. Let $m^{+}=\sum_{i=1}^{n}d^{+}(i)$ and $m^{-}=\sum_{i=1}^{n}d^{-}(i)$. Now, we are ready to define the general modularity $Q$ as below
\begin{align}\label{GQ}
Q=Q^{+}-Q^{-},
\end{align}
where
\begin{align*}
Q^{+}=\begin{cases}
\sum_{i,j}(A^{+}(i,j)-\frac{d^{+}(i)d^{+}(j)}{2m^{+}})\delta(\hat{\ell}(i),\hat{\ell}(j)), & \mbox{when~} m^{+}>0,\\
0, & \mbox{when~} m^{+}=0,
\end{cases},
\end{align*}
\begin{align*}
Q^{-}=\begin{cases}
\sum_{i,j}|(A^{-}(i,j)-\frac{d^{-}(i)d^{-}(j)}{2m^{-}})\delta(\hat{\ell}(i),\hat{\ell}(j))|, & \mbox{when~} m^{-}>0,\\
0, & \mbox{when~} m^{-}=0,
\end{cases},
\end{align*}
$\hat{\ell}$ is the $n\times1$ vector such that $\hat{\ell}(i)$ denotes the cluster that node $i$ belongs to, and
\begin{align*}
\delta(\hat{\ell}(i),\hat{\ell}(i))=\begin{cases}
1& \mbox{when~} \hat{\ell}(i)=\hat{\ell}(j),\\
0, & \mbox{otherwise}.
\end{cases},
\end{align*}
For weighted network in which all entries of $A$ are nonnegative (i.e, $Q=Q^{+}$ since $Q^{-}=0$ when $A(i,j)\geq 0$ for all $i,j$), $Q$ reduces to the classical Newman's modularity \cite{newman2006modularity}. For weighted network in which $A$ contains negative entries, $Q$ is an extension of classical modularity by considering negative entries of $A$. The intuition of designing $Q^{-}$ by summarizing absolute values of $(A^{-}(i,j)-\frac{d^{-}(i)d^{-}(j)}{2m^{-}})\delta(\hat{\ell}_{i},\hat{\ell}_{j})$ comes from the fact that a negative $A(i,j)$ may not mean that nodes $i$ and $j$ tend to be in different cluster since $A$'s negative elements may be generated from Normal distribution, Logistic distribution or some other distributions. We set $Q=Q^{+}-Q^{-}$ empirically since such modularity is a good measure to investigate the performance of different algorithms.

In Eq (\ref{GQ}), we write $Q$ as $Q(\bullet)$  for convenience where $\bullet$ denotes certain community detection method since $\hat{\ell}$ is obtained by running the community detection method $\bullet$ to $A$ with $K$ communities. Next, we define the effectiveness of the general modularity $Q$. Since $\hat{f}$ is stronger criterion than the Hamming error \cite{SCORE}, for numerical studies, the Hamming error rate defined below is applied to investigate performances of algorithms.
\begin{align*}
\hat{f}(\bullet)=n^{-1}\mathrm{min}_{J\in\mathcal{P}_{K}}\|\hat{Z}J-Z\|_{0},
\end{align*}
where $\mathcal{P}_{K}$ is the set of all $K\times K$ permutation matrices, the matrix $\hat{Z}\in\mathbb{R}^{n\times K}$ is defined as $\hat{Z}(i,k)=1$ if $\hat{\ell}(i)=k$ and 0 otherwise for $i\in[n],k\in[K]$, and $\hat{\ell}$ is the label vector returned from applying method $\bullet$ to $A$ with $K$ communities. Sure, $\hat{f}(\bullet1)<\hat{f}(\bullet2)$ means method $\bullet1$ outperforms method $\bullet2$. Now, we are ready to define the effectiveness of $Q$ when $\hat{f}(\bullet1)\neq \hat{f}(\bullet2)$:
\begin{align*}
E_{Q}(\bullet1,\bullet2)=\begin{cases}
1& \mbox{when~} (\hat{f}(\bullet1)-\hat{f}(\bullet2))(Q(\bullet2)-Q(\bullet1))>0,\\
-1, & \mbox{otherwise},
\end{cases}.
\end{align*}
where we do not consider the case $\hat{f}(\bullet1)=\hat{f}(\bullet2)$ because $Q(\bullet2)=Q(\bullet1)$ for this case and it does not tell the effectiveness of $Q$. On the one hand, $E_{Q}(\bullet1,\bullet2)=1$ means that if method $\bullet1$ outperforms method $\bullet2$ (i.e., $\hat{f}(\bullet1)<\hat{f}(\bullet2)$), then we also have $Q(\bullet1)\geq Q(\bullet2)$, i.e., the generality modularity $Q$ is effective. On the other hand, $E_{Q}(\bullet1,\bullet2)=-1$ means that if method $\bullet1$ outperforms method $\bullet2$, we have $Q(\bullet1)<Q(\bullet2)$ which means that $Q$ is ineffective. For any experiment, suppose we generate $N$ adjacency matrices $A$ under a community detection model, we obtain $N$ numbers of $E_{Q}(\bullet1,\bullet2)$. The ratio of effectiveness is defined as
\begin{align}\label{RationEffectiveness}
R_{E_{Q}(\bullet1,\bullet2)}=\frac{\sum_{E_{Q}(\bullet1,\bullet2)=1}E_{Q}(\bullet1,\bullet2)}{N-N_{0}},
\end{align}
where $N_{0}$ is the number of adjacency matrices such that $\hat{f}(\bullet1)=\hat{f}(\bullet2)$ since the effectiveness of the generality modularity is defined when $\hat{f}(\bullet1)\neq\hat{f}(\bullet2)$. Sure, a lager $R_{E_{Q}(\bullet1,\bullet2)}$ indicates the effectiveness of the generality modularity $Q$ obtained by applying methods $\bullet1$ and $\bullet2$.
\subsection*{Simulations}
In numerical simulations, we aim at comparing nDFA with DFA under DCDFM by reporting $\hat{f}(nDFA)$ and $\hat{f}(DFA)$, and investigating the effectiveness of Q by reporting $R_{E_{Q}(DFA,nDFA)}$ computed from nDFA and DFA.

In all simulated data, unless specified, set $n=400, K=4$, generate $\ell$ such that node belongs to each community with equal probability, and let $\rho>0$ be a parameter such that $\theta(i)=\rho\times\mathrm{rand}(1)$, where $\mathrm{rand}(1)$ is a random value in the interval $(0,1)$. $\rho$ is regarded as sparsity parameter controlling the sparsity of network $\mathcal{N}$.  When $P, Z, \Theta$ are set, $\Omega$ is $\Omega=\Theta ZPZ'\Theta$.  Generate the symmetric adjacency matrix $A$ by letting $A(i,j)$ generated from a distribution $\mathcal{F}$ with expectation $\Omega(i,j)$. Different distributions will be studied in simulations, and we show the error rates of different methods, averaged over 100 random runs for each setting of some model parameters.
\subsubsection*{Experiment 1: Normal distribution}
This experiment studies the case when $\mathcal{F}$ is Normal distribution. Set $P$ as
\[P=\begin{bmatrix}
    -1&-0.4&0.5&0.2\\
    -0.4&0.9&0.2&-0.2\\
    0.5&0.2&0.8&0.3\\
    0.2&-0.2&0.3&-0.9
\end{bmatrix}.\]
Since $\mathcal{F}$ is Normal distribution, elements of $P$ are allowed to be negative under DCDFM. Generate the symmetric adjacency matrix $A$ by letting $A(i,j)$ be a random variable generated from $\mathrm{Normal}(\Omega(i,j),\sigma^{2}_{A})$ for some $\sigma^{2}_{A}$. Note that the only criteria for choosing $P$ is, $P$ should satisfy Eq (\ref{DefinMM}), and elements of $P$ should be positive or negative depending on distribution $\mathcal{F}$. See, if $\mathcal{F}$ is Normal distribution as in this experiment, $P$ can have negative entries; if $\mathcal{F}$ is Bernoulli or Poisson or Binomial distribution, all entries of $P$ should be nonnegative.

\begin{figure}
\centering
\subfigure[Normal distribution]{\includegraphics[width=0.33\textwidth]{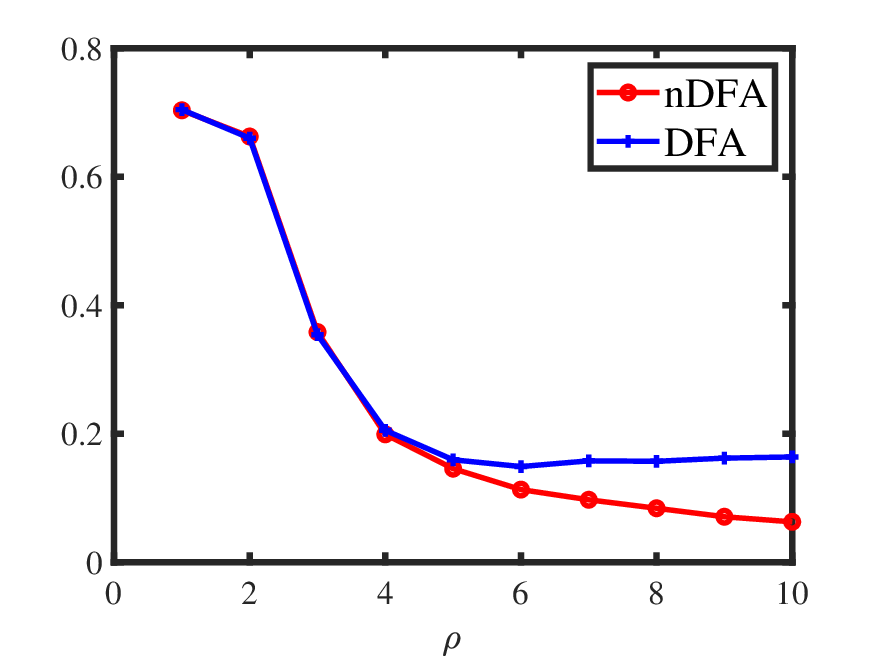}}
\subfigure[Normal distribution]{\includegraphics[width=0.33\textwidth]{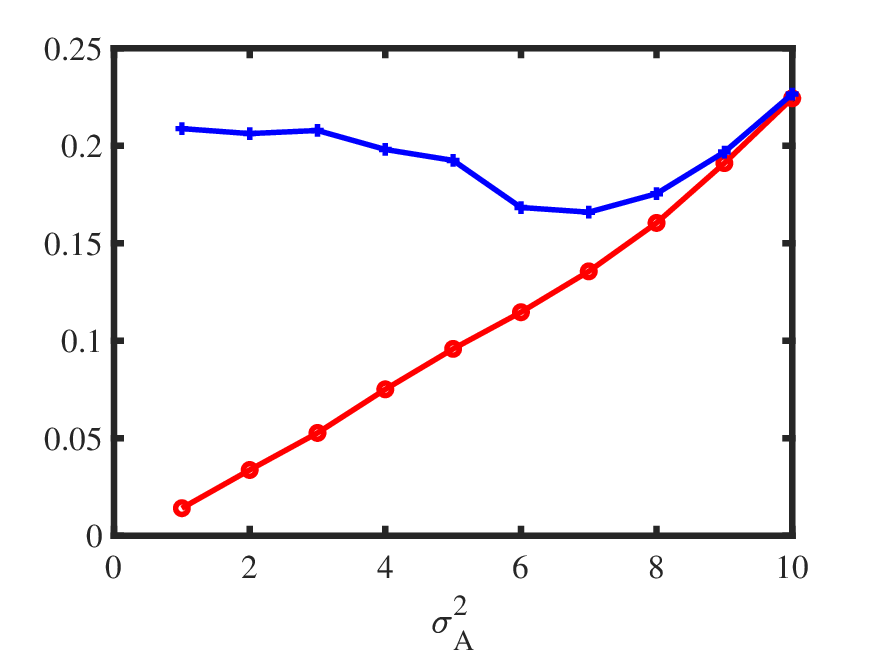}}
\subfigure[Binomial distribution]{\includegraphics[width=0.33\textwidth]{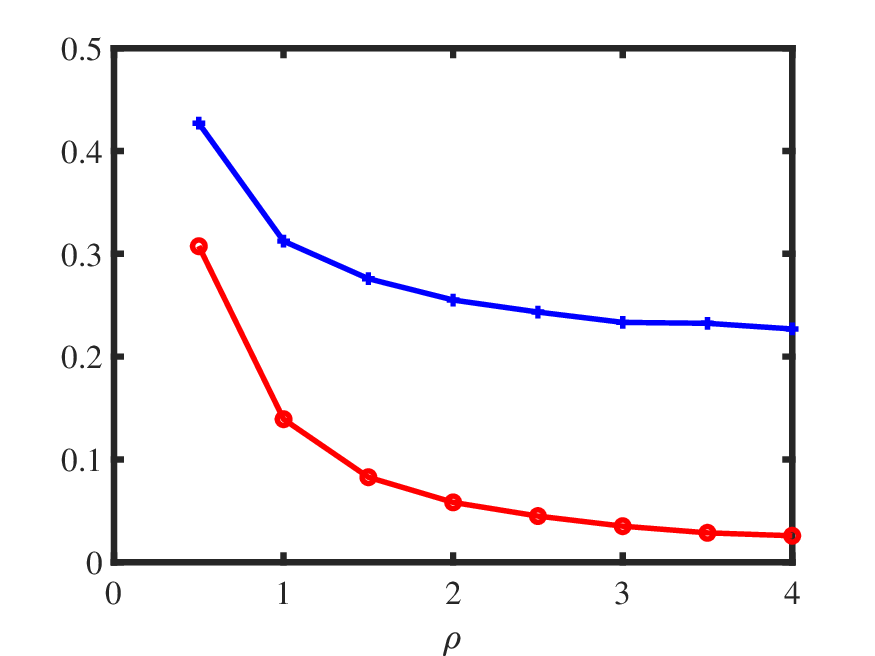}}
\subfigure[Binomial distribution]{\includegraphics[width=0.33\textwidth]{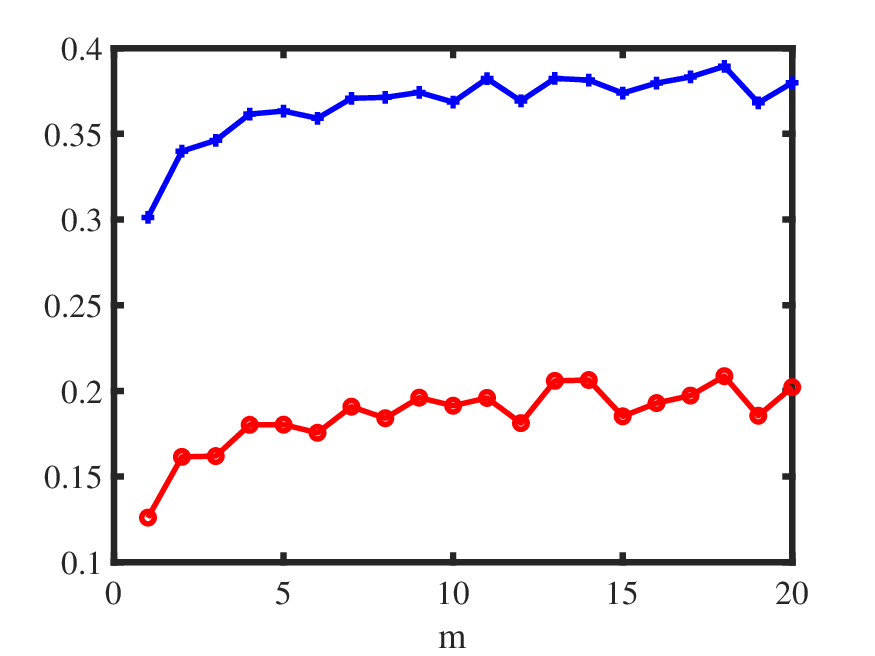}}
\subfigure[Bernoulli distribution]{\includegraphics[width=0.33\textwidth]{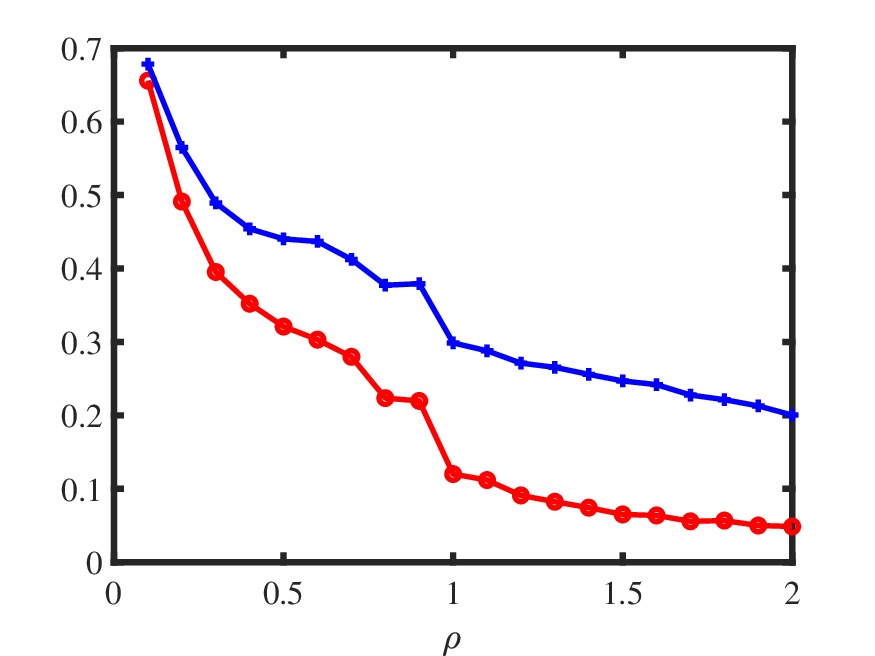}}
\subfigure[Poisson distribution]{\includegraphics[width=0.33\textwidth]{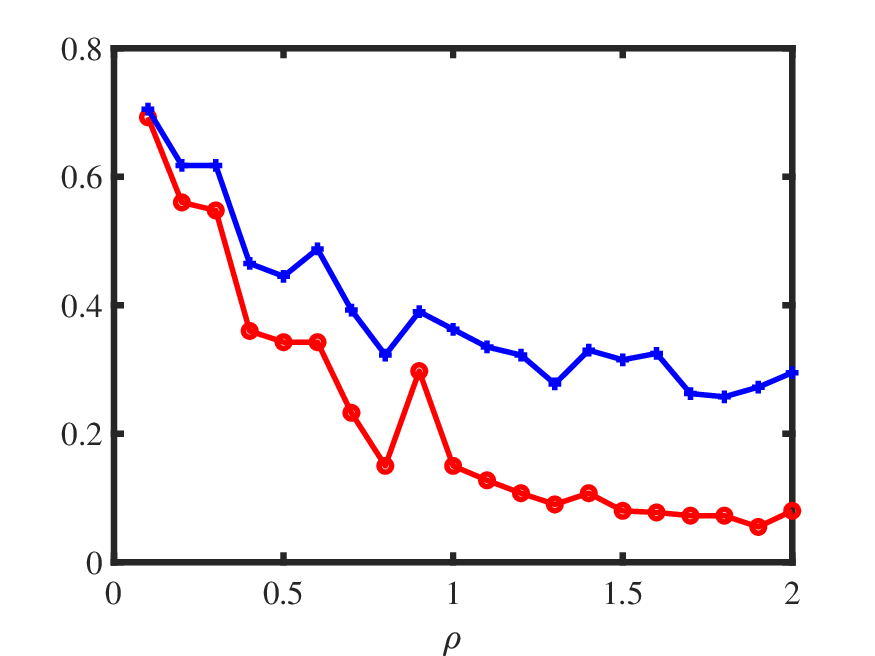}}
\subfigure[Logistic distribution]{\includegraphics[width=0.33\textwidth]{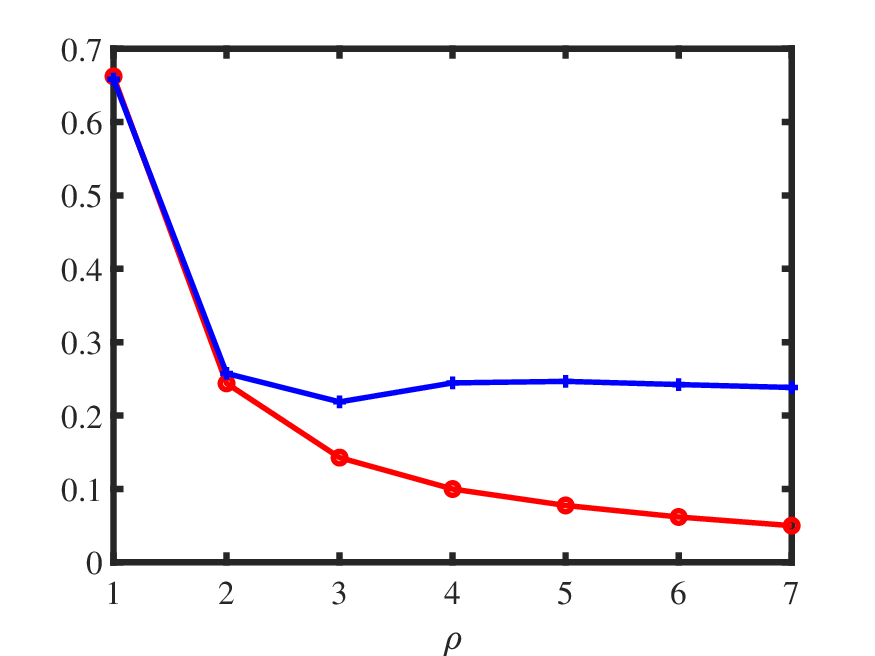}}
\subfigure[Logistic distribution]{\includegraphics[width=0.33\textwidth]{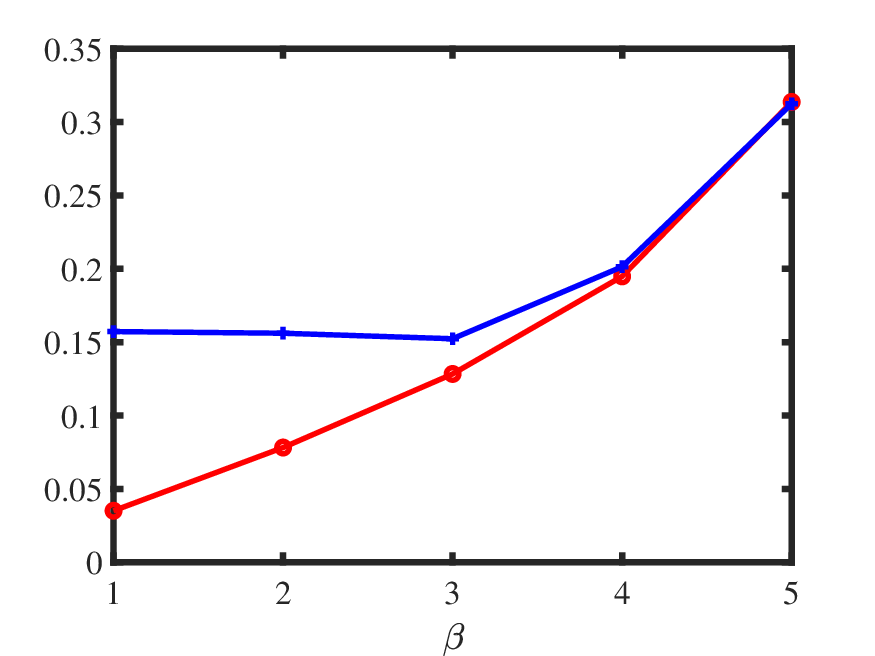}}
\subfigure[Signed network]{\includegraphics[width=0.33\textwidth]{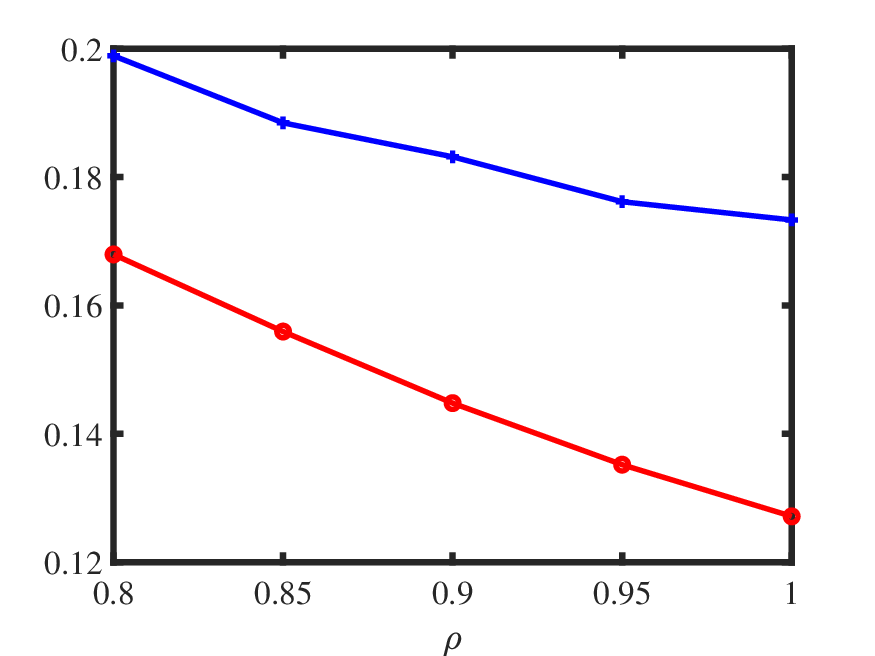}}
\caption{Numerical results of Experiments 1-6. y-axis: error rate.}
\label{EX} 
\end{figure}

\texttt{Experiment 1[a]: Changing $\rho$.} Let $\sigma^{2}_{A}=4$, and $\rho$ range in $\{1,2,\ldots,10\}$. In panel (a) of Figure \ref{EX}, we plot the error against $\rho$. For larger $\rho$, we get denser networks, and the two methods perform better. When $\rho$ is larger than 5, nDFA outperforms DFA. Meanwhile, Experiment 1[a] generates totally $10\times 100=1000$ adjacency matrices, where $10$ is the cardinality of $\{1,2,\ldots,10\}$, and 100 is the repetition for each $\rho$. Among the 1000 adjacency matrices, we calculate $R_{E_{Q}(DFA,nDFA)}$ based on DFA and nDFA. $R_{E_{Q}(DFA,nDFA)}$ for Experiment 1[a] is reported in Table \ref{SimulatedQ}, and we see that $R_{E_{Q}(DFA,nDFA)}$ is $82.59\%$ (a value much larger than 50\%), suggesting the effectiveness of the general modularity.  Similar illustrations on the calculation of $R_{E_{Q}(DFA,nDFA)}$ hold for other simulated experiments in this paper.

\texttt{Experiment 1[b]: Changing $\sigma^{2}_{A}$.} Let $\rho=10$, and $\sigma^{2}_{A}$ range in $\{1,2,\ldots,10\}$. In panel (b) of Figure \ref{EX}, we plot the error against $\sigma^{2}_{A}$. For larger $\sigma^{2}_{A}$, theoretical upper bound of $\hat{f}$ is larger for nDFA as shown by the first bullet given after Corollary \ref{cor}. Thus, the increasing error of nDFA when $\sigma^{2}_{A}$ increases is consistent with our theoretical findings. Meanwhile, the numerical results also tell us that nDFA significantly outperforms DFA since DFA always  performs poor when there exists node heterogeneity for each node.
\subsubsection*{Experiment 2: Binomial distribution}
This experiment considers the case when $\mathcal{F}$ is Binomial distribution. For binomial distribution, as discussed in \emph{Example} \ref{Binomial}, $P$ should be an nonnegative matrix. Set $P$ as
\[P=\begin{bmatrix}
    1&0.4&0.5&0.2\\
    0.4&0.9&0.2&0.2\\
    0.5&0.2&0.8&0.3\\
    0.2&0.2&0.3&0.9
\end{bmatrix}.\]
Since $\Omega=\Theta ZPZ'\Theta$, generate the symmetric matrix $A$ such that $A(i,j)$ is a random variable generated according to $\mathrm{Binomial}(m,\frac{\Omega(i,j)}{m})$ for some positive integer $m$.

\texttt{Experiment 2[a]: Changing $\rho$.} Let $m=5$, and $\rho$ range in $\{0.5,1,\ldots,4\}$. Note that since $\frac{\Omega(i,j)}{m}$ is a probability and $\Omega(i,j)\leq \rho$, $\rho$ should be set lesser than $m$. In panel (c) of Figure \ref{EX}, we plot the error against $\rho$. We see that the two methods perform better as $\rho$ increases and nDFA behaves much better than DFA.

\texttt{Experiment 2[b]: Changing $m$.} Let $\rho=1$, and $m$ range in $\{1,2,\ldots,20\}$. In panel (d) of Figure \ref{EX}, we plot the error against $m$. For larger $m$, both two methods perform poorer, and this phenomenon occurs because $A(i,j)$ may take more integers as $m$ increases when $\mathcal{F}$ is Binomial distribution. The results also show that nDFA performs much better than DFA  when considering variation of node degree.
\begin{remark}\label{SimulatedAvisual}
For visuality, we plot $A$ generated under DFM when $\mathcal{F}$ is Binomial distribution. Let $n=24,K=2$. Let $Z(i,1)=1$ for $1\leq i\leq 12$, $Z(i,2)=1$ for $13\leq i\leq24$. Let $m=5$, and $\Theta=0.7I$ (i.e., a DFM case). Set $P$ as
\[P=\begin{bmatrix}
    1&0.3\\
    0.3&0.9
\end{bmatrix}.\]
For above setting, two different adjacency matrices are generated under DFM in Figure \ref{BinomialA} where we also report error rates for DFA and nDFA. Meanwhile, since $A$ and $Z$ are known here, one can run DFA and nDFA directly to $A$ in Figure \ref{BinomialA} with two communities to check the error rates of DFA and nDFA. Furthermore, we also plot adjacency matrices for Bernoulli distribution, Poisson distribution and Signed network.
\begin{figure}
\centering
\subfigure[]{\includegraphics[width=0.49\textwidth]{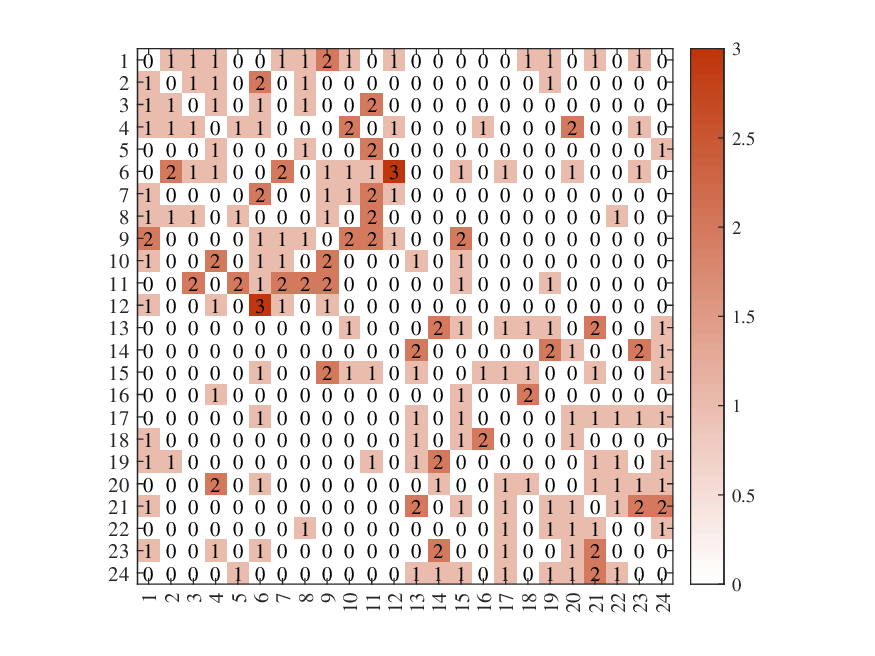}}
\subfigure[]{\includegraphics[width=0.49\textwidth]{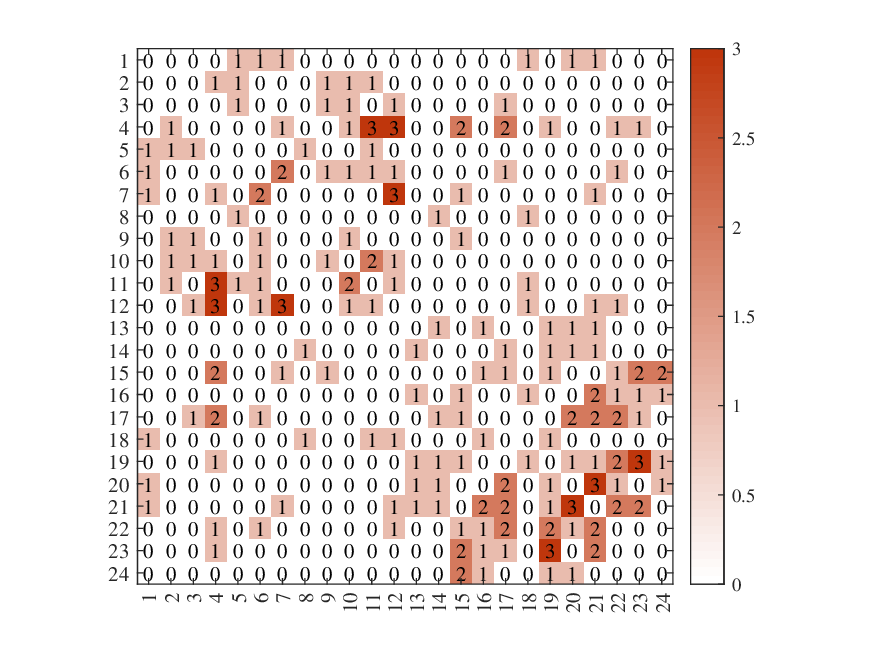}}
\caption{For adjacency matrix in panel (a),  $\hat{f}(DFA)=0.0833, \hat{f}(nDFA)=0$. For adjacency matrix in panel (b), $\hat{f}(DFA)=0.0417, \hat{f}(nDFA)=0.1250$.  For both panels, $R_{E_{Q}(DFA,nDFA)}=1$. x-axis: row nodes; y-axis: column nodes.}
\label{BinomialA} 
\end{figure}
\end{remark}
\subsubsection*{Experiment 3: Bernoulli distribution}
In this experiment, let $\mathcal{F}$ be Bernoulli distribution such that $A(i,j)$ is random variable generated from $\mathrm{Bernoulli}(\Omega(i,j))$. Set $P$ same as Experiment 2.

\texttt{Experiment 3: Changing $\rho$.} Let $\rho$ range in $\{0.1,0.2,\ldots,2\}$. In panel (e) of Figure \ref{EX}, we plot the error against $\rho$. Similar as Experiment 2[a], nDFA outperforms DFA.
\begin{remark}
For visuality, we plot $A$ generated under DFM when $\mathcal{F}$ is Bernoulli. Let $n,K,Z,\Theta,P$ be the same as Remark \ref{SimulatedAvisual}. Two different adjacency matrices shown in Figure \ref{BernoulliA} are generated under above setting.
\begin{figure}
\centering
\subfigure[]{\includegraphics[width=0.49\textwidth]{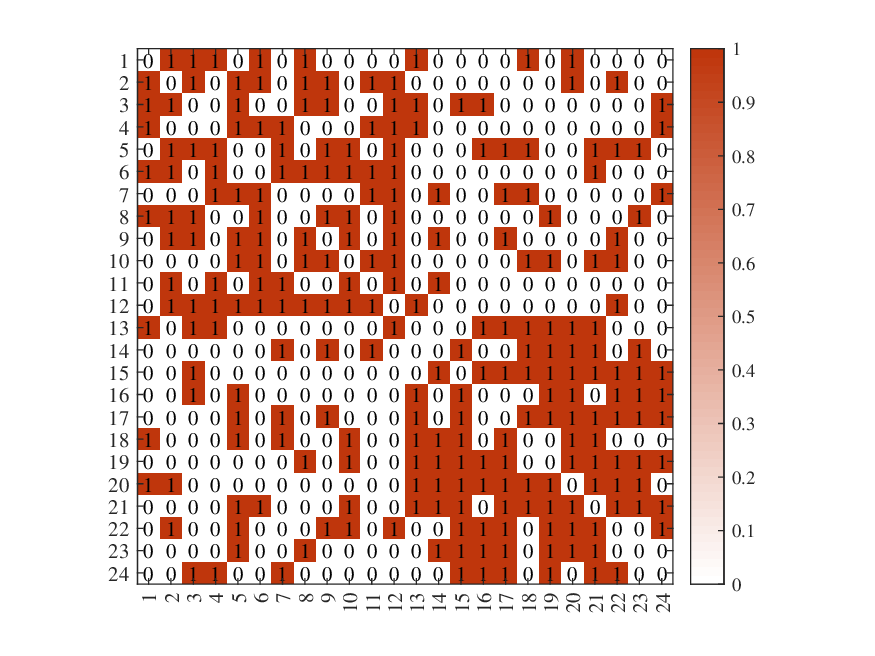}}
\subfigure[]{\includegraphics[width=0.49\textwidth]{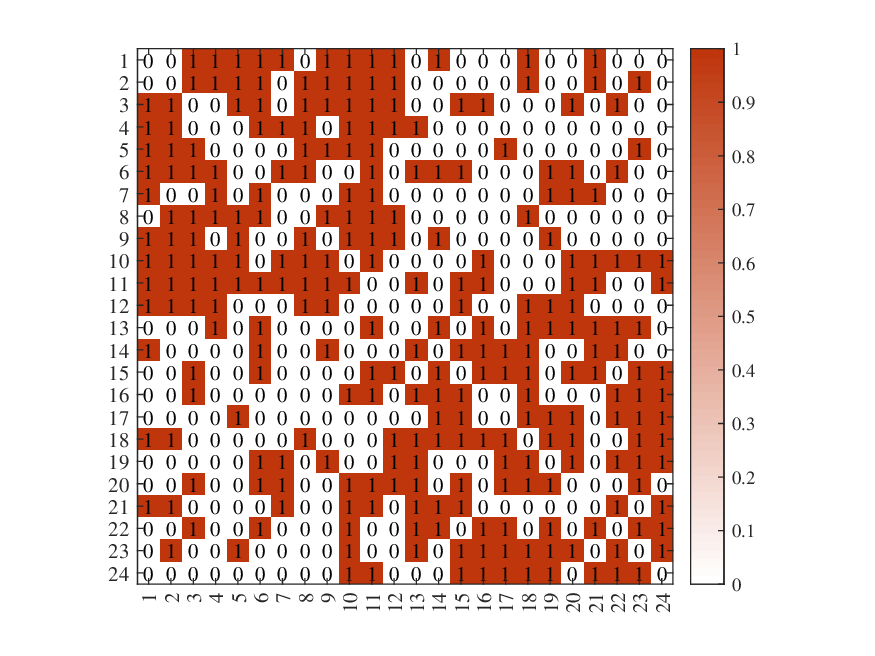}}
\caption{For both adjacency matrices in panels (a) and (b), error rates for DFA and nDFA are 0. x-axis: row nodes; y-axis: column nodes.}
\label{BernoulliA} 
\end{figure}
\end{remark}
\subsubsection*{Experiment 4: Poisson distribution}
This experiment focuses on the case when $\mathcal{F}$ is Poisson distribution such that $A(i,j)$ is random variable generated from $\mathrm{Poisson}(\Omega(i,j))$. Set $P$ same as Experiment 2.

\texttt{Experiment 4: Changing $\rho$.} Let $\rho$ range in $\{0.1,0.2,\ldots,2\}$. In panel (f) of Figure \ref{EX}, we plot the error against $\rho$. The results are similar as that of Experiment 2[a], and nDFA enjoys better performance than DFA.
\begin{remark}
For visuality, we plot $A$ generated under DFM when $\mathcal{F}$ is Poisson. $n,K,Z,\Theta,P$ are set the same as Remark \ref{SimulatedAvisual}. Two different adjacency matrices shown in Figure \ref{PoissonA} are generated under above setting.
\begin{figure}
\centering
\subfigure[]{\includegraphics[width=0.49\textwidth]{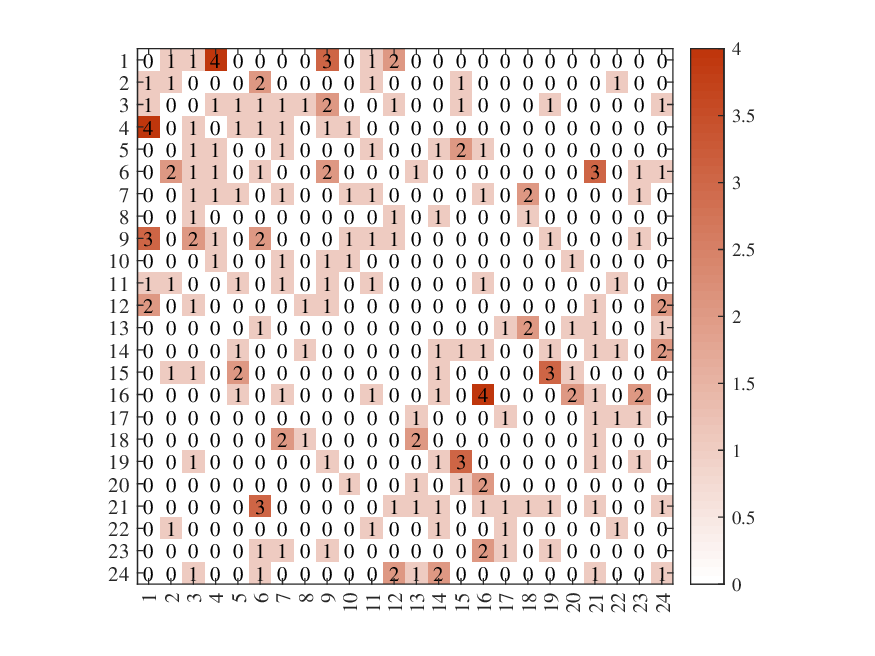}}
\subfigure[]{\includegraphics[width=0.49\textwidth]{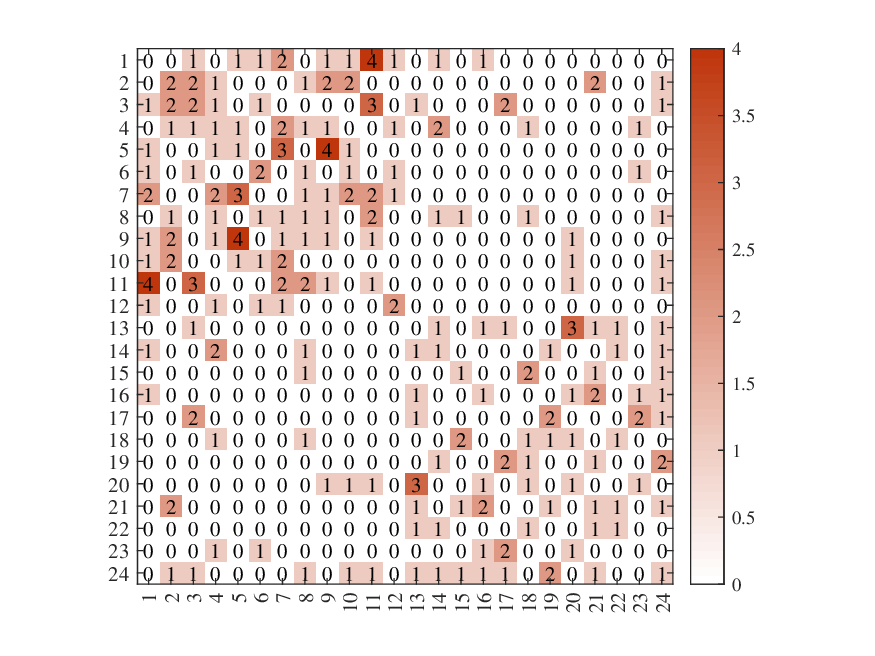}}
\caption{For adjacency matrix in panel (a),  $\hat{f}(DFA)=0.2500, \hat{f}(nDFA)=0.1250$. For adjacency matrix in panel (b), $\hat{f}(DFA)=0.1250, \hat{f}(nDFA)=0.0833$.  For both panels, $R_{E_{Q}(DFA,nDFA)}=1$. x-axis: row nodes; y-axis: column nodes.}
\label{PoissonA} 
\end{figure}
\end{remark}
\subsubsection*{Experiment 5: Logistic distribution}
In this experiment, let $\mathcal{F}$ be Logistic distribution such that $A(i,j)$ is random variable generated from $\mathrm{Logistic}(\Omega(i,j),\beta)$. Set $P$ same as Experiment 1.

\texttt{Experiment 5[a]: Changing $\rho$.} Let $\beta=1$, and  $\rho$ range in $\{1,2,\ldots,7\}$. In panel (g) of Figure \ref{EX}, we plot the error against $\rho$. The results are similar as that of Experiment 1[a], and nDFA outperforms DFA.

\texttt{Experiment 5[b]: Changing $\beta$.} Let $\rho=7$, and  $\beta$ range in $\{1,2,\ldots,5\}$. Panel (h) of Figure \ref{EX} plots the error against $\rho$. The results are similar as that of Experiment 1[b], and nDFA outperforms DFA.
\subsubsection*{Experiment 6: Signed network}
In this experiment, let $\mathbb{P}(A(i,j)=1)=\frac{1+\Omega(i,j)}{2}$ and $\mathbb{P}(A(i,j)=-1)=\frac{1-\Omega(i,j)}{2}$ such that all elements of $A$ are either $1$ or $-1$. Set $P$ same as Experiment 1.

\texttt{Experiment 6: Changing $\rho$.} Let $n=1600$, and  $\rho$ range in $\{0.8,0.85,0.9,0.95,1\}$. In panel (i) of Figure \ref{EX}, we plot the error against $\rho$. And nDFA outperforms DFA.
\begin{remark}
For visuality, we plot $A$ generated under DFM when $\mathbb{P}(A(i,j)=1)=\frac{1+\Omega(i,j)}{2}$ and $\mathbb{P}(A(i,j)=-1)=\frac{1-\Omega(i,j)}{2}$ for signed network. $n,K,Z,\Theta$ are set the same as Remark \ref{SimulatedAvisual}, and $P$ is set as
\[P=\begin{bmatrix}
    -1&0.3\\
    0.3&0.9
\end{bmatrix}.\]
Two different adjacency matrices generated under above setting are shown in Figure \ref{SignedA}.
\begin{figure}
\centering
\subfigure[]{\includegraphics[width=0.49\textwidth]{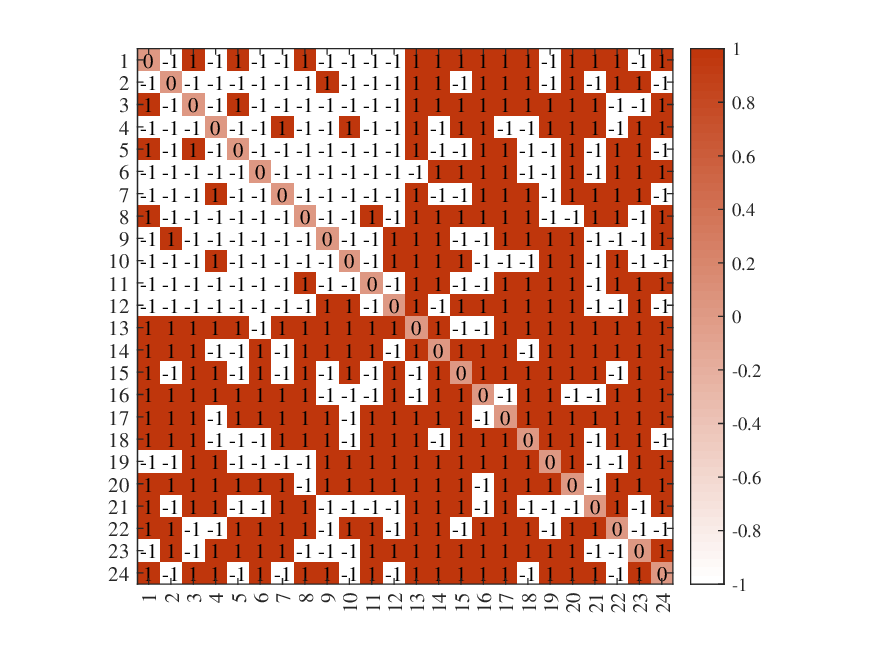}}
\subfigure[]{\includegraphics[width=0.49\textwidth]{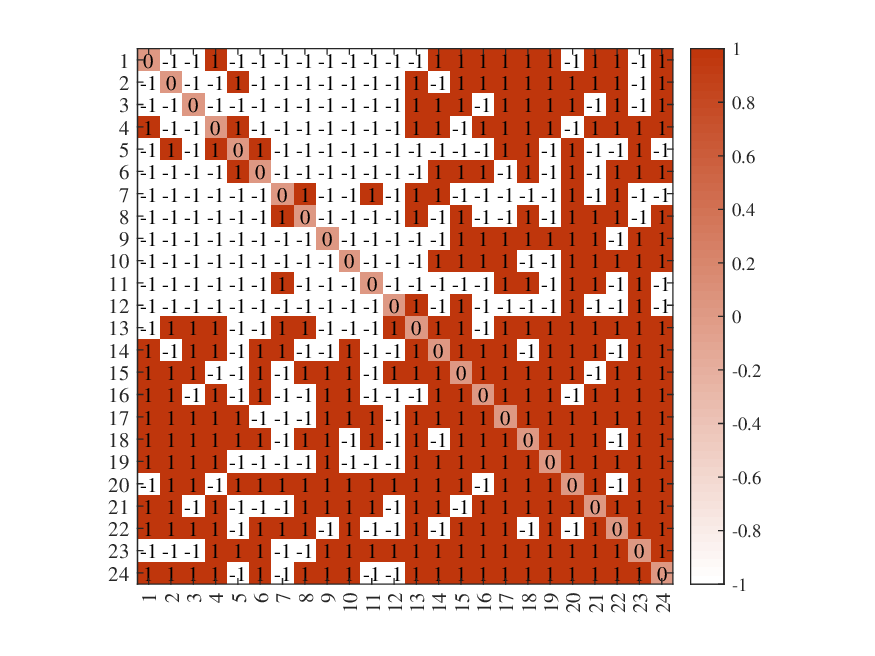}}
\caption{For adjacency matrix in panel (a),  $\hat{f}(DFA)=0.0833, \hat{f}(nDFA)=0.0417$. For adjacency matrix in panel (b), $\hat{f}(DFA)=0, \hat{f}(nDFA)=0$.  For panel (a), $R_{E_{Q}(DFA,nDFA)}=1$. x-axis: row nodes; y-axis: column nodes.}
\label{SignedA} 
\end{figure}
\end{remark}
\begin{table}[t]
\caption{$R_{E_{Q}(DFA,nDFA)}$ for Experiments 1-6, where Ex denotes Experiment. For these experiments, values of $R_{E_{Q}(DFA,nDFA)}$ are larger than 80\%, suggesting the effectiveness of the general modularity.}
\label{SimulatedQ}
\vskip 0.15in
\begin{center}
\begin{small}
\begin{tabular}{lcccccccccr}
\toprule
&Ex~1[a]&Ex~1[b]&Ex~2[a]&Ex~2[b]&Ex~3&Ex~4&Ex~5[a]&Ex~5[b]&Ex~6\\
\midrule
$R_{E_{Q}(DFA,nDFA)}$&82.59\%&93.69\%&99\%&99.8\%&92.32\%&95.74\%&89.99\%&83.64\%&100\%\\
\bottomrule
\end{tabular}
\end{small}
\end{center}
\vskip -0.1in
\end{table}
\subsection*{Real data}
In real data analysis, instead of simply using our general modularity for comparative analysis, we also consider the topological comparative evaluation framework proposed in \cite{orman2012comparative}. We only consider two topological approaches embeddedness which measures how much the direct neighbours of a node belong to its own
community and community size which is an important characteristic of the community structure \cite{orman2012comparative}, because the internal transitivity, scaled density, average distance and hub dominance introduced in \cite{orman2012comparative} only work for un-weighted networks while we will consider weighted networks in this part. Now we provide the definition of embeddedness \cite{lancichinetti2010characterizing,orman2012comparative}: for node $i$, let $d_{\mathrm{int}}(i)=\sum_{j:\hat{\ell}(j)=\hat{\ell}(i)}A(i,j)$ be the internal degree of node $i$ belonging to cluster $\hat{\ell}(i)$ and $d(i)=\sum_{j=1}^{n}A(i,j)$ be the total degree of node $i$, where $\hat{\ell}$ is the estimated nodes labels for certain method $\bullet$. The embeddedness of node $i$ is defined as
\begin{align*}
e(i)=\frac{d_{\mathrm{int}}(i)}{d(i)}\equiv\frac{\sum_{j:\hat{\ell}(j)=\hat{\ell}(i)}A(i,j)}{\sum_{j=1}^{n}A(i,j)},
\end{align*}
where this definition of embeddedness extends that of \cite{lancichinetti2010characterizing,orman2012comparative} from un-weighted network to weighted network whose adjacency matrix is connected and has nonnegative entries. Extending the definition of embeddedness for adjacency matrix in which there may exist negative elements is an interesting problem, and we leave it for our future work. Meanwhile, $e(i)$ is only defined for one node $i$, to capture embeddedness for all nodes, we define the overall embedbedness (OE for short) depending on method $\bullet$ as
\begin{align*}
OE(\bullet)=\frac{\sum_{i=1}^{n}e(i)}{n}.
\end{align*}
As analyzed in \cite{orman2012comparative}, the maximal $OE(\bullet)$ of 1 is reached when all the neighbours are in its community for all nodes (i.e., $d_{\mathrm{int}}(i)=d(i)$ for all $i$). However, if method $\bullet$ puts all nodes  (or a majority of nodes) into one community, then it can also make $OE(\bullet)$ equal to 1 (or close to 1). Therefore, simply using the overall embedbedness to compare the performances of different community detection methods is not enough, we need to consider the general modularity $Q(\bullet)$ and community size. Set
\begin{align*}
\tau(\bullet)=\frac{\mathrm{max}_{1\leq k\leq K}|i:\hat{\ell}(i)=k|}{n},
\end{align*}
where $\tau(\bullet)$ measures how much the size of the largest estimated cluster to the network size. If $\tau(\bullet)$ is 1 (or close to 1), it means that method $\bullet$ puts all nodes (or a majority of nodes) into one community. For real-world networks with known true labels, we let $error(\bullet)$ denotes the error rate of method $\bullet$. Finally, $T(\bullet)$ denote the run-time of method $\bullet$. For real-world networks analyzed in this paper, we will report the general modularity $Q$, the overall embeddedness $OE$, community size parameter $\tau$, error rate $error$ (for real-world network with known true labels) and run-time $T$ of nDFA and DFA for our comparative analysis.
\subsubsection*{Real-world un-weighted networks}
In this section, four real-world un-weighted networks with known labels are studied to investigate nDFA's empirical performance. Some basic information of the four data are displayed in Table \ref{real4}, where Karate, Dolphins, Polbooks and Weblogs are short for Zachary's karate club, Dolphin social network, Books about US politics and Political blogs, and the four datasets can be downloaded from
\url{http://www-personal.umich.edu/~mejn/netdata/}. For these real-world un-weighted networks, their true labels are suggested by the original authors, and they are regarded as the ``ground truth''. Brief introductions of the four networks can be found in \cite{DCSBM,RSC,SCORE,jin2021improvements}, and reference therein. Similar as the real data study part in \cite{qing2021DFM}, since all entries of adjacency matrices of the four real data sets are $1$ or $0$ (i.e., the original adjacency matrices of the four real data are un-weighted), to construct weighted networks, we assume there exists noise such that we have the observed matrix $\hat{A}$ at hand where $\hat{A}=A+W$ with the noise matrix $W\sim\mathrm{Normal}(0,\sigma^{2}_{W})$, i.e, use $\hat{A}$ as input matrix in nDFA and DFA instead of using $A$. We let $\sigma^{2}_{W}$ range in $\{0,0.01,0.02,\ldots,0.2\}$. For each $\sigma^{2}_{W}$, we report error rate of different methods averaged over 50 random runs and aim to study nDFA's behaviors when $\sigma^{2}_{W}$ increase. Note that similar as in the perturbation analysis of \cite{qing2021DFM}, we can add a noise matrix $W$ whose entries have mean 0 and finite variance in our theoretical analysis of nDFA, and we do not consider perturbation analysis during our theoretical study of nDFA for convenience in this paper. We only consider the influence of noise matrix in our numerical study part to reveal the performance stability of our algorithm nDFA.
\begin{table}[t]
\caption{Four real-world un-weighted social networks with known label information, where $d_{\mathrm{min}}=\mathrm{min}_{i\in[n]}\sum_{j=1}^{n}
A(i,j)$ is the minimum degree, $d_{\mathrm{max}}$ is the maximum degree.}
\label{real4}
\vskip 0.15in
\begin{center}
\begin{small}
\begin{tabular}{lcccr}
\toprule
\#&Karate&Dolphins&Polbooks&Weblogs\\
\midrule
$n$&34&62&92&1222\\
$K$&2&2&2&2\\
$d_{\mathrm{min}}$&1&1&1&1\\
$d_{\mathrm{max}}$&17&12&24&351\\
\bottomrule
\end{tabular}
\end{small}
\end{center}
\vskip -0.1in
\end{table}
\begin{table}[t]
\caption{The general modularity, overall embeddedness, community size parameter, error rate and run-time of DFA and nDFA for networks in Table \ref{real4}.}
\label{Qreal4}
\vskip 0.15in
\begin{center}
\begin{small}
\begin{tabular}{lcccr}
\toprule
&Karate&Dolphins&Polbooks&Weblogs\\
\midrule
$Q(\mathrm{DFA})$&32.8590&36.9214&183.7647&4279.0987\\
$OE(\mathrm{DFA})$&0.9009&0.8462&0.9562&0.7389\\
$error(\mathrm{DFA})$&0/34&12/62&4/92&437/1222\\
$\tau(\mathrm{DFA})$&18/34&53/62&47/92&1071/1222\\
$T(\mathrm{DFA})$&0.05s&0.07s&0.08s&0.14s\\
\hline
$Q(\mathrm{nDFA})$&32.8590&64.5818&184.7767&7142.2134\\
$OE(\mathrm{nDFA})$&0.9009&0.9565&0.9706&0.9405\\
$error(\mathrm{nDFA})$&0/34&1/62&3/92&64/1222\\
$\tau(\mathrm{nDFA})$&18/34&40/62&46/92&670/1222\\
$T(\mathrm{nDFA})$&0.05s&0.04s&0.05s&0.09s\\
\bottomrule
\end{tabular}
\end{small}
\end{center}
\vskip -0.1in
\end{table}
\begin{figure}
\centering
\subfigure[Karate]{\includegraphics[width=0.245\textwidth]{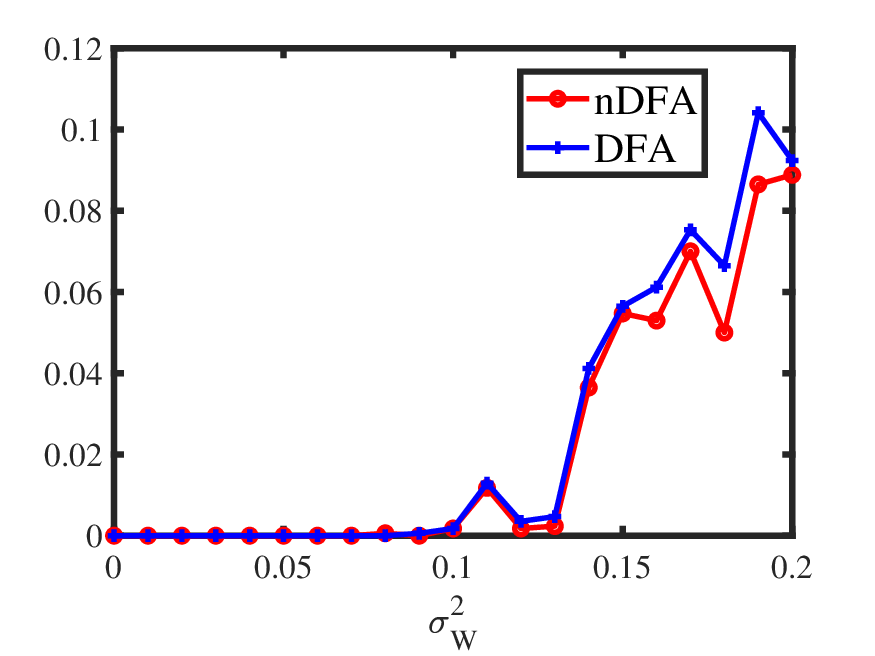}}
\subfigure[Dolphins]{\includegraphics[width=0.245\textwidth]{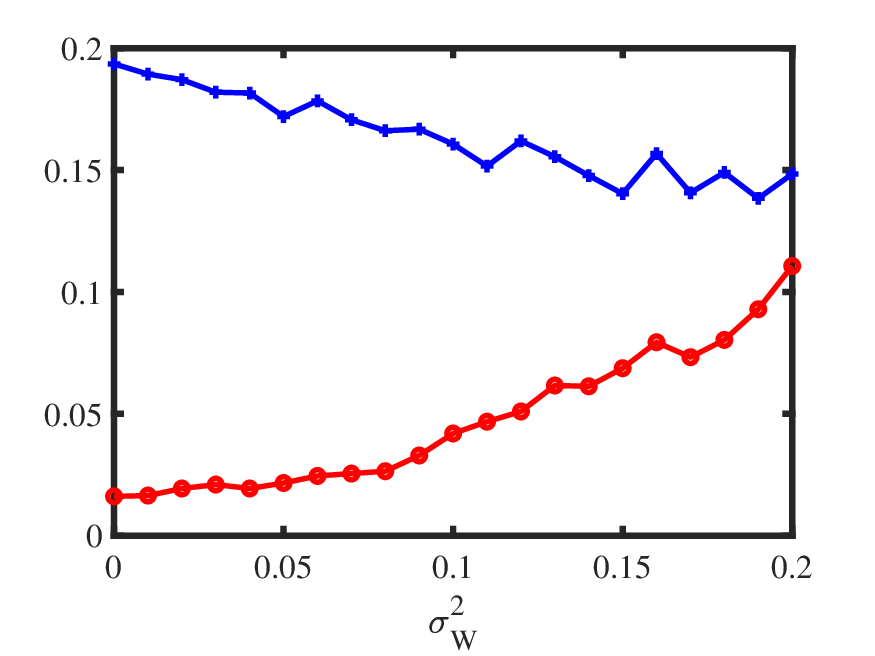}}
\subfigure[Polbooks]{\includegraphics[width=0.245\textwidth]{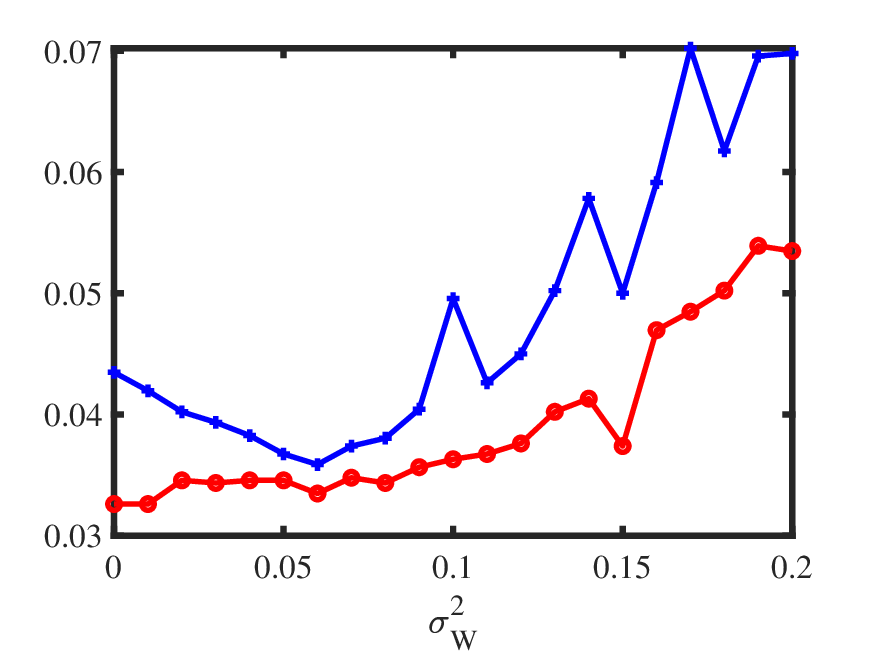}}
\subfigure[Weblogs]{\includegraphics[width=0.245\textwidth]{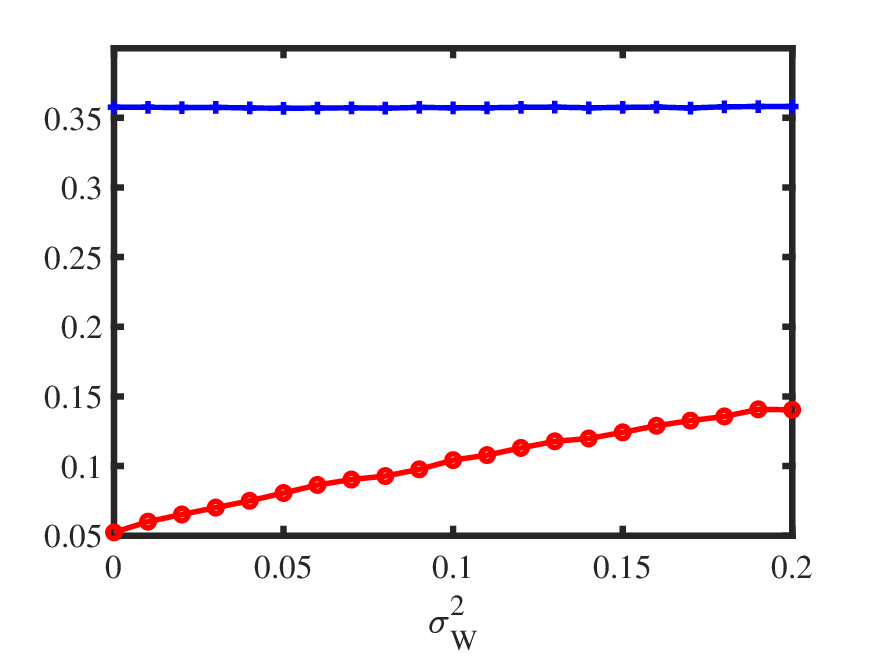}}
\caption{Error rates of nDFA and DFA on the four real-world networks. y-axis: error rate.}
\label{Real} 
\end{figure}

Figure \ref{Real} displays the error rates against $\sigma^{2}_{W}$ for the four real-world social networks. When noise matrix $W$ has small variance, nDFA has stable performances. When elements of $W$ varies significantly, nDFA's error rates increases. DFA also has stable performances when $\sigma^{2}_{W}$ is small, except that DFA always performs poor on Dolphins and Weblogs networks even for the case that there is no noise ($\sigma^{2}_{W}=0$ means a case without noise). For the two networks Karate and Polbooks, nDFA has similar performances as DFA and both methods enjoy satisfactory performances. For Dolphins and Weblogs, nDFA performs much better than DFA. Especially, for Weblogs network, DFA's error rates are always around $35\%$, which is a large error rate, while nDFA's error rates are always lesser than $15\%$ even for a noise matrix with large variance. This can be explained by the fact that the node degree in Weblogs network varies heavily, as analyzed in \cite{DCSBM, SCORE}. Since nDFA is designed under DCDFM considering node heterogeneity while DFA is designed under DFM without considering node heterogeneity, naturally, nDFA can enjoy better performances than DFA on real-world networks with variation in node degree.

Meanwhile, $Q,OE,error,\tau$ and $T$ obtained by applying DFA and nDFA to adjacency matrices $A$ for the above four real-world networks with known nodes labels are reported in Table \ref{Qreal4}. Combine results in Table \ref{Qreal4} and Figure  \ref{Real}, we see that when nDFA has smaller error rates than DFA, nDFA has larger modularity than DFA, and this suggests the general modularity is effective for un-weighted networks (note that, the general modularity is exactly the Newman's modularity when all entries of $A$ are nonnegative). For Dolphins, Polbooks and Weblogs, we see that both the overall embeddedness and modularity of nDFA are larger than that of DFA, which suggests that nDFA returns more accurate estimation on nodes labels than DFA, and this is consistent with the fact that nDFA has smaller error rates than DFA for these three networks. Compared with nDFA whose error rates are small, $\tau$ of DFA for Dolphins and Weblogs are much larger than that of nDFA, which suggests that DFA tends to put nodes into one community. Meanwhile, small error rates, large overall embeddedness (close to 1), and medium size of the largest estimated community of nDFA suggest that these four networks enjoy nice community structure for community detection. Sure, both methods run fast on these four networks.
\subsubsection*{Real-world weighted networks}
\begin{figure}
\centering
\subfigure[Karate-weighted]{\includegraphics[width=0.49\textwidth]{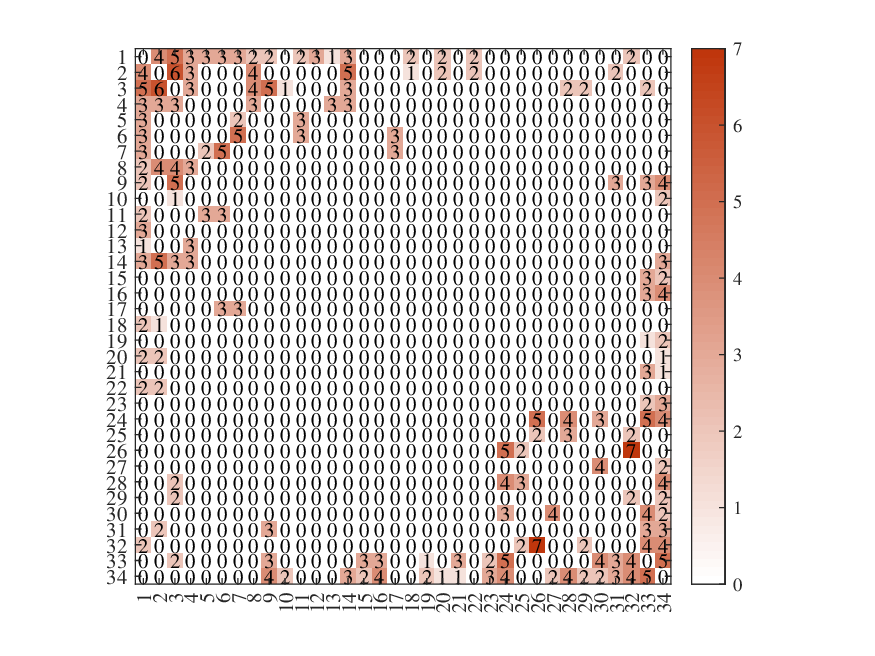}}
\subfigure[Gahuku-Gama subtribes]{\includegraphics[width=0.49\textwidth]{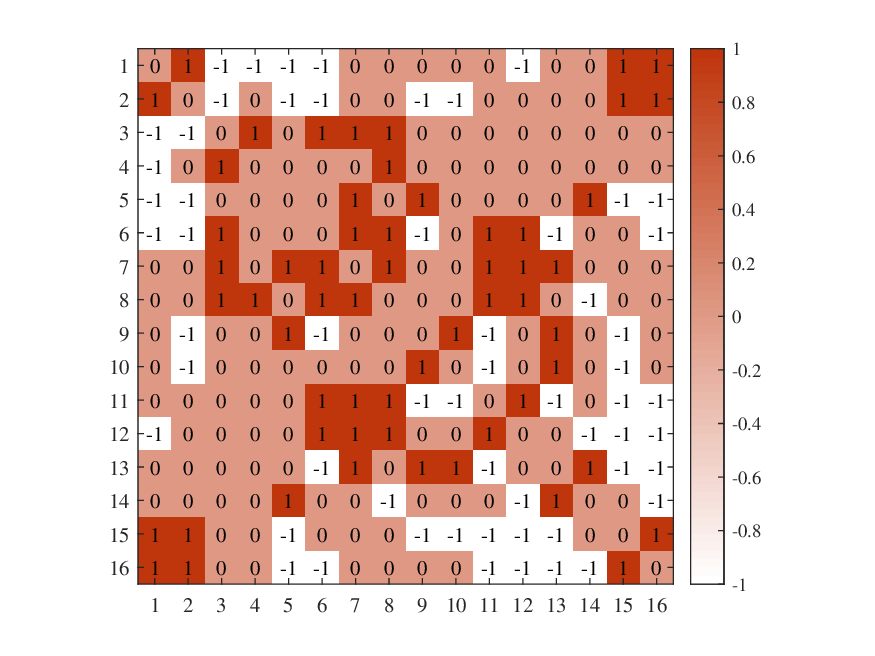}}
\caption{Adjacency matrices of Karate-weighted and Gahuku-Gama subtribes.}
\label{RealA} 
\end{figure}

In this section, we apply nDFA and DFA to five real-world weighted networks Karate club weighted network (Karate-weighted for short), Gahuku-Gama subtribes network, the Coauthorships in network science network (CoauthorshipsNet for short), Condensed matter collaborations 1999 (Con-mat-1999 for short) and  Condensed matter collaborations 2003 (Con-mat-2003 for short). For visualization, Figure \ref{RealA} shows adjacency matrices of the first two weighted networks. Table \ref{realdata5} summaries basic information for the five networks. Detailed information of the five networks can be found below.

\emph{Karate-weighted}: This weighted network is collected from a university karate club. In this weighted network, node denotes member, and edge between two nodes indicates the relative strength of the associations. Actually, this network is the weighted version of Karate club network. So, the number of communities is 2 and true labels for all members are known for Karate-weighted. This data can be downloaded from \url{http://vlado.fmf.uni-lj.si/pub/networks/data/ucinet/ucidata.htm#kazalo}.

\emph{Gahuku-Gama subtribes}: This data is the signed social network of tribes of the Gahuku–Gama alliance structure of the Eastern Central Highlands of New Guinea. This network has 16 tribes, and  positive or negative link between two tribes means
they are allies or enmities, respectively. Meanwhile, there are 3 communities in this network, and we use nodes labels shown in Figure 9 (b) from \cite{yang2007community} as ground truth. This data can be downloaded from \url{http://konect.cc/}(see also \cite{kunegis2013konect}). Note that since the overall embeddedness is defined for adjacency matrix with nonnegative entries, it is not applicable for this network.

\begin{figure}
\centering
\subfigure[CoauthorshipsNet1589]{\includegraphics[width=0.245\textwidth]{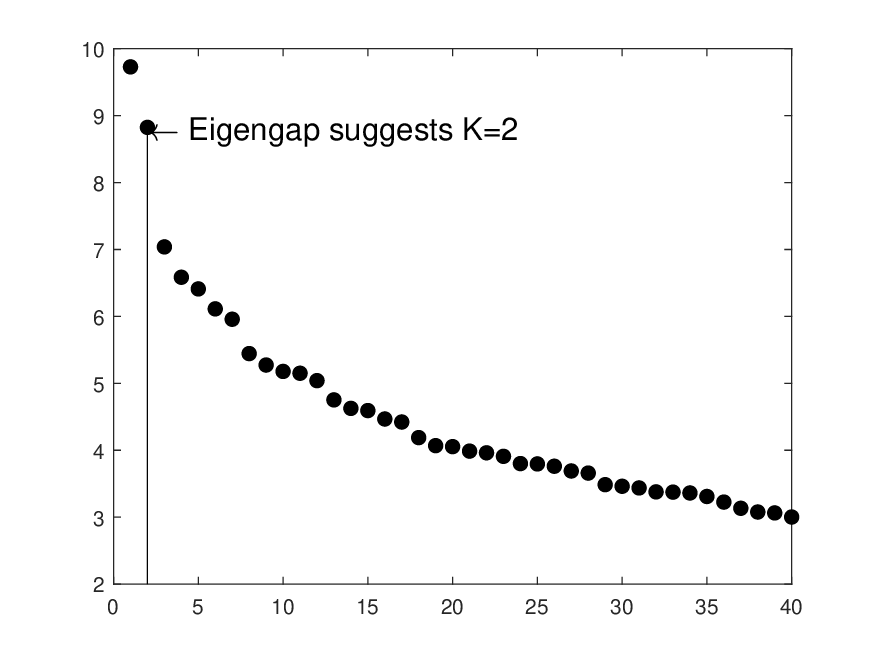}}
\subfigure[CoauthorshipsNet379]{\includegraphics[width=0.245\textwidth]{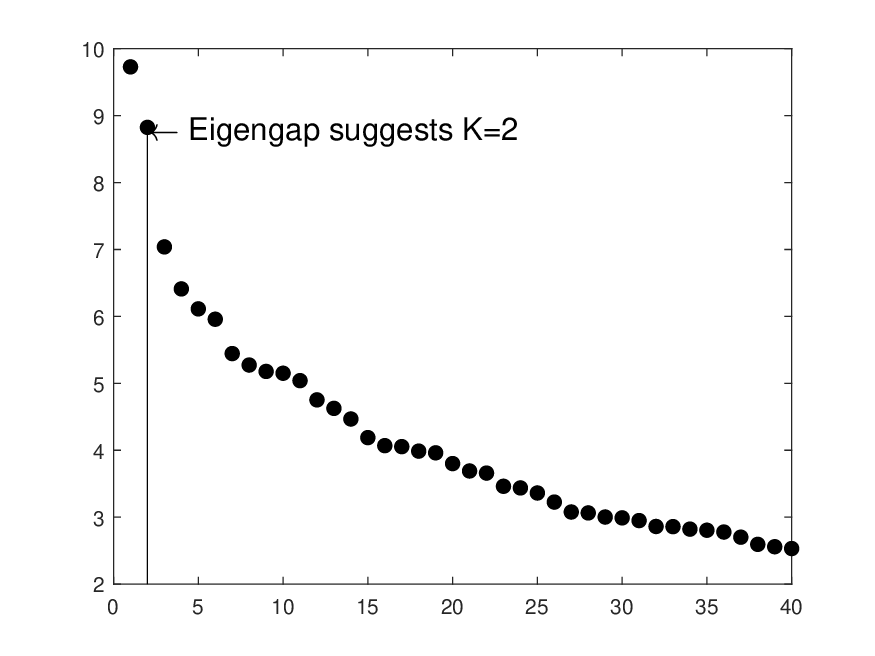}}
\subfigure[Con-mat-1999]{\includegraphics[width=0.245\textwidth]{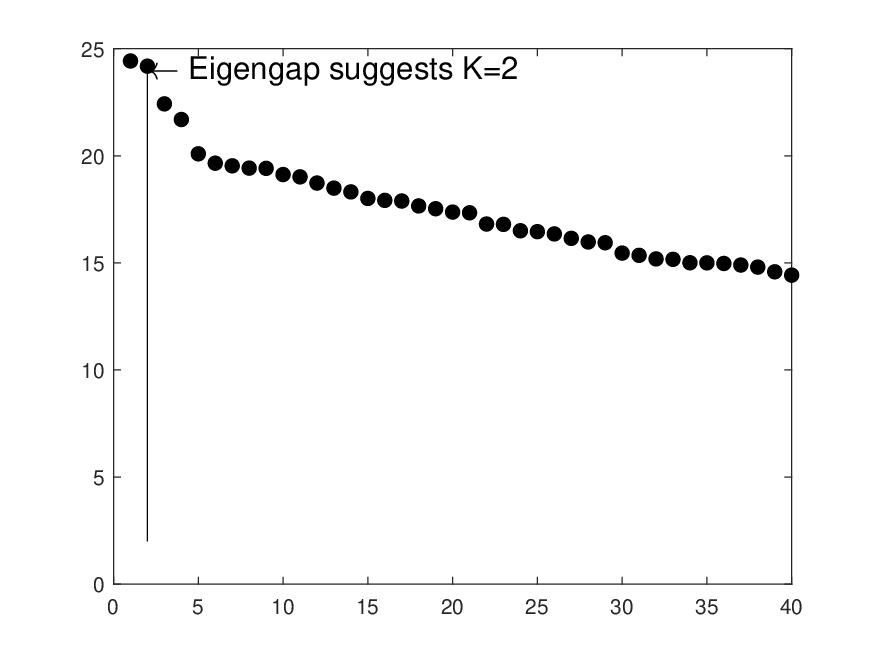}}
\subfigure[Con-mat-2003]{\includegraphics[width=0.245\textwidth]{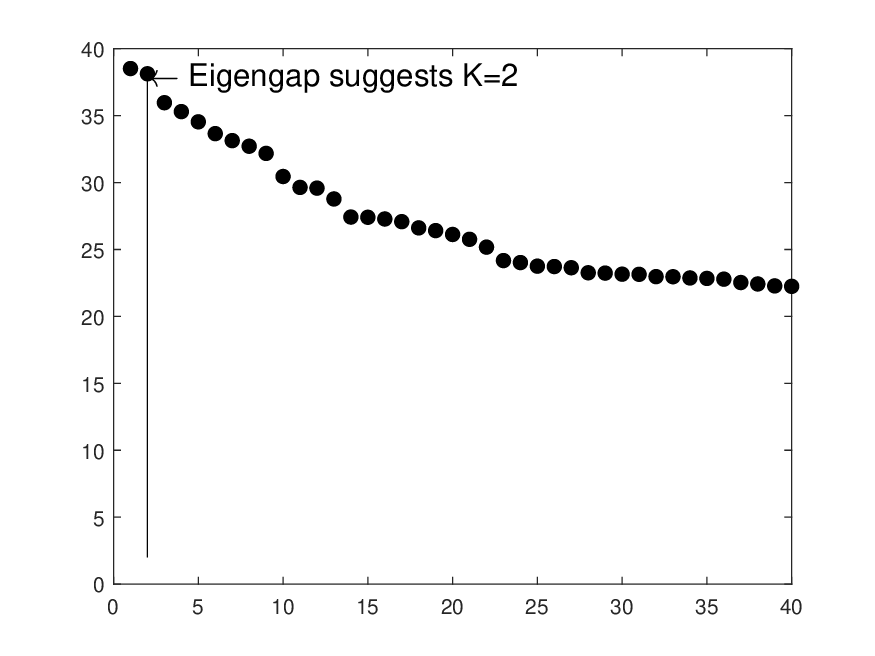}}
\caption{Panels (a)-(d) show leading 40 eigenvalues of adjacency matrices for CoauthorshipsNet1589, CoauthorshipsNet379, Con-mat-1999 and Con-mat-2003, respectively.}
\label{CoNetK} 
\end{figure}

\emph{CoauthorshipsNet}: This data can be downloaded from
\url{http://www-personal.umich.edu/~mejn/netdata/}.
In CoauthorshipsNet, node means scientist and weights mean coauthorship, where weights are assigned by the original papers. For this network, there is no ground truth about nodes labels, and the numbers of communities are unknown. The CoauthorshipsNet has 1589 nodes, however its adjacency matrix is disconnected. Among the 1589 nodes, there are totally 396 disconnected components, and only 379 nodes fall in the largest connected component. For convenience, we use CoauthorshipsNet1589 to denote the original network, and CoauthorshipsNet379 to denote the giant component. To find the number of communities for CoauthorshipsNet, we plot the leading 40 eigenvalues of their adjacency matrices. Results shown in Figure \ref{CoNetK} suggest that the number of communities is 2, where \cite{rohe2016co} also applies the idea of eigengap to estimate the number of communities for real-world networks. Note that though CoauthorshipsNet1589 is disconnected, we can still apply nDFA and DFA on it since there is no requirement on network connectivity when applying DFA and nDFA. Note that since the overall embeddedness is defined for adjacency matrix that is connected, it is not applicable for CoauthorshipsNet1589.

\emph{Con-mat-1999}: This data can be downloaded from
\url{http://www-personal.umich.edu/~mejn/netdata/}. In this network, node denotes scientists and edge weights are provided by the original papers. The largest connected component for this data has 13861 nodes. Figure \ref{CoNetK} suggests $K=2$ for this data.

\emph{Con-mat-2003}: It is updated network of Con-mat-1999 and the largest connected component has 27519 nodes. Figure \ref{CoNetK} suggests $K=2$ for Con-mat-2003.

\begin{table}[h!]
\footnotesize
	\centering
	\caption{Summary information for real-world weighted networks used in this paper.}
	\label{realdata5}
\begin{tabular}{cccccccccc}
\hline\hline
&Source&$n$&$K$&$\mathrm{max}_{i,j}A(i,j)$&$\mathrm{min}_{i,j}A(i,j)$&\#Edges&\%Positive edges\\
\hline
Karate-weighted&\cite{zachary1977information}&34&2&7&0&78&100\%\\
Gahuku-Gama subtribes&\cite{read1954cultures}&16&3&1&-1&58&50\%\\
CoauthorshipsNet1589&\cite{newman2006finding}&1589&Unknown&4.75&0&2742&100\%\\
CoauthorshipsNet379&\cite{newman2006finding}&379&Unknown&4.75&0&914&100\%\\
Con-mat-1999&\cite{newman2001structure}&13861&Unknown&22.3333&0&44619&100\%\\
Con-mat-2003&\cite{newman2001structure}&27519&Unknown&35.2&0&116181&100\%\\
\hline\hline
\end{tabular}
\end{table}

\begin{table}[htp]
\caption{$Q,OE,\tau$ and $T$ of DFA and nDFA for networks in Table \ref{realdata5}.}
\label{Qreal5}
\vskip 0.15in
\begin{center}
\begin{small}
\begin{tabular}{lcccccr}
\toprule
&Karate-weighted&Gahuku-Gama subtribes&CoauthorshipsNet1589&CoauthorshipsNet379&Con-mat-1999&Con-mat-2003\\
\midrule
$Q(\mathrm{DFA})$&105.0433&9.7931&28.8493&28.1765&49.7391&111.0945\\
$OE(\mathrm{DFA})$&0.9250&-&-&0.9611&0.9997&0.9995\\
$\tau(\mathrm{DFA})$&18/34&7/16&1585/1589&375/379&13850/13861&27516/27519\\
$T(\mathrm{DFA})$&0.05s&0.05s&0.132s&0.065s&11.81s&47.51s\\
\hline
$Q(\mathrm{nDFA})$&105.0433&9.7931&238.8005&220.2177&7104.9&15641\\
$OE(\mathrm{nDFA})$&0.9250&-&-&0.9857&0.9420&0.9174\\
$\tau(\mathrm{nDFA})$&18/34&7/16&1025/1589&257/379&10483/13861&20915/27519\\
$T(\mathrm{nDFA})$&0.05s&0.05s&0.183s&0.051s&11.97s&47.23s\\
\bottomrule
\end{tabular}
\end{small}
\end{center}
\vskip -0.1in
\end{table}

\begin{figure}
\centering
\subfigure[]{\includegraphics[width=0.49\textwidth]{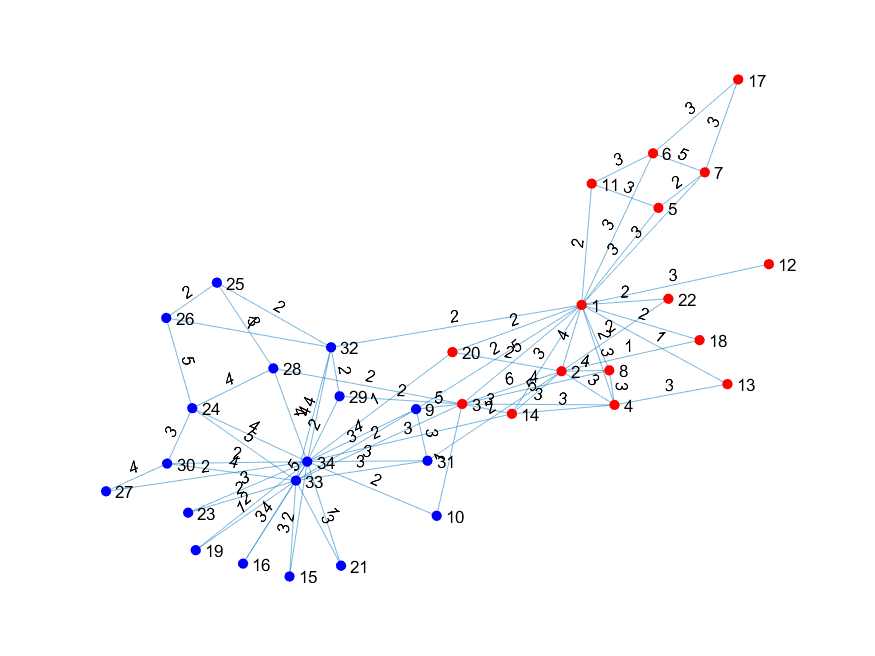}}
\subfigure[]{\includegraphics[width=0.49\textwidth]{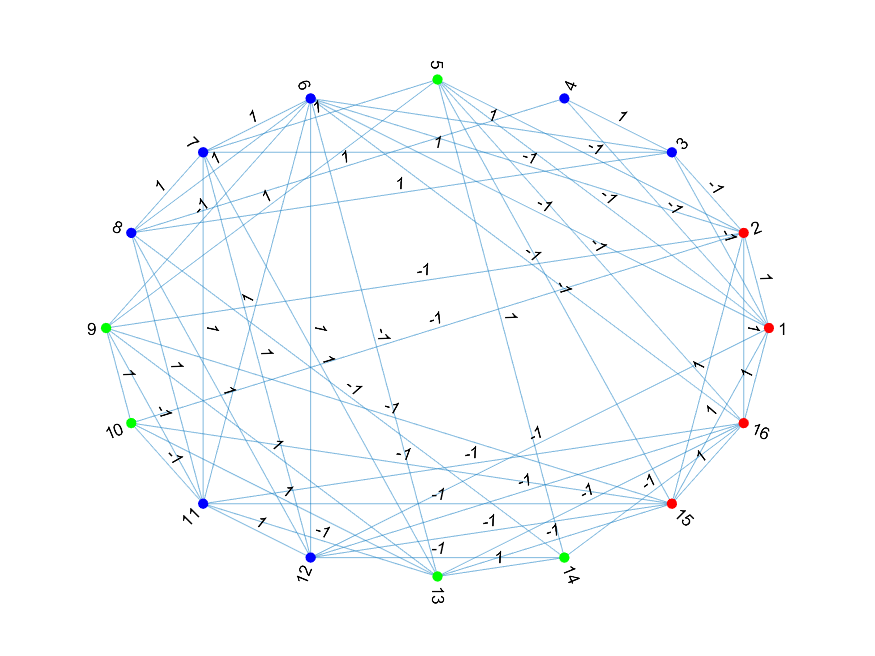}}
\caption{Panel (a) and panel (b) show nDFA's detection results of Karate-weighted and Gahuku-Gama subtribes, respectively. In this two panels, different colors are used to distinguish different communities.}
\label{RandC} 
\end{figure}

\begin{figure}
\centering
\subfigure[]{\includegraphics[width=0.49\textwidth]{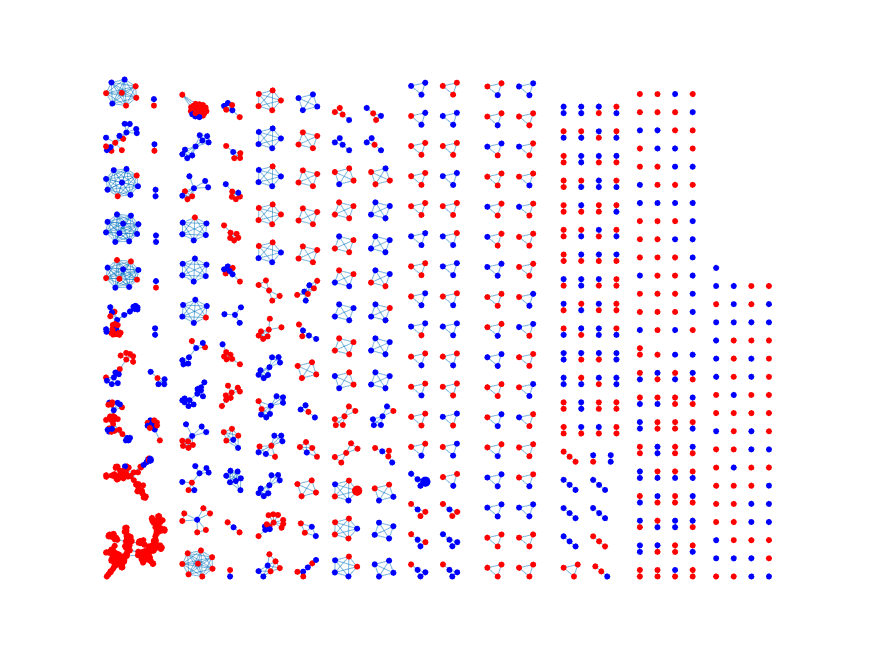}}
\subfigure[]{\includegraphics[width=0.49\textwidth]{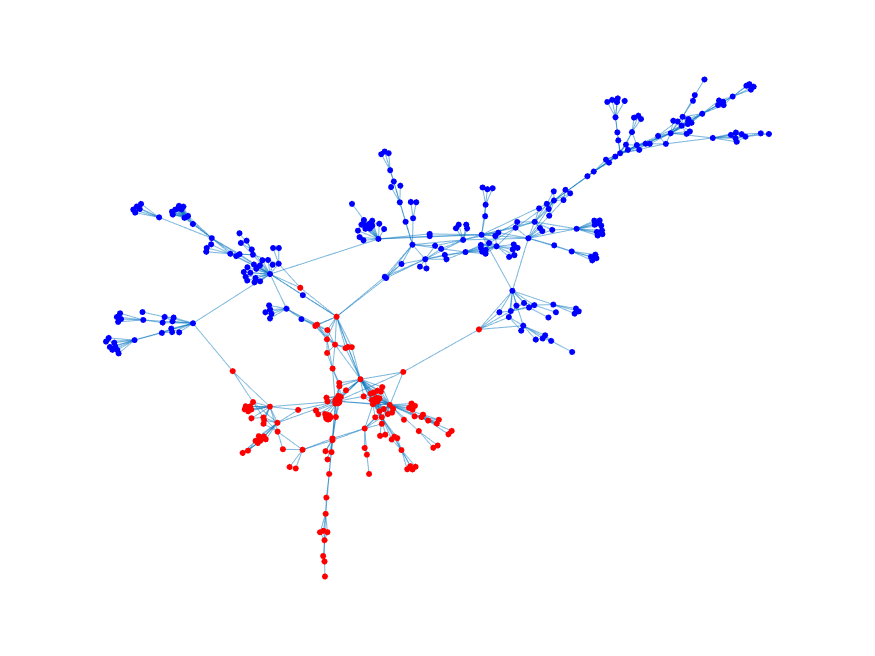}}
\caption{Panel (a) and panel (b) show nDFA's detection results of CoauthorshipsNet1589 and CoauthorshipsNet379, respectively. Here, different colors are used to distinguish different communities. For visualization, two nodes are connected by a line if there is a positive edge weight between them, and we do not show edge weights here.}
\label{NetCoNet} 
\end{figure}

\begin{figure}
\centering
\includegraphics[width=0.66\textwidth]{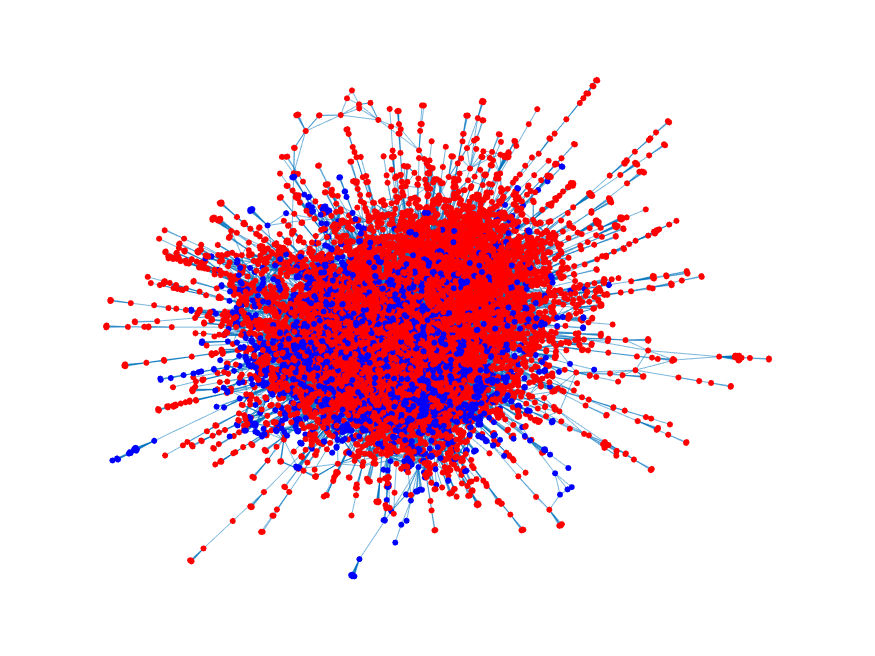}
\caption{nDFA's detection results of Cont-mat-1999. Colors indicate communities. For visualization, two nodes are connected by a line if there is a positive edge weight between them, and we do not show edge weights here.}
\label{NetCon} 
\end{figure}

We apply nDFA and DFA on Karate-weighted and Gahuku-Gama subtribes, and find that error rates for both methods on both data are zero, suggesting that nDFA and DFA perform perfect on this two networks. For visualization, Figures \ref{RandC}, \ref{NetCoNet} and \ref{NetCon} show community detection results by applying nDFA on these weighted networks except Con-mat-2003 whose size is too large to plot using the graph command of MATLAB. Note that disconnected components and isolated nodes can also be classified by nDFA as shown in panel (a) of Figure \ref{NetCoNet}, and this guarantees the widely applicability of nDFA since it can deal with disconnected weighted network even with isolated nodes. Table \ref{Qreal5} records $Q,OE,\tau$ and $T$ for the five weighted networks, and we find that $Q(\mathrm{nDFA})$ is much larger than $Q(\mathrm{DFA})$ for CoauthorshipsNet, Con-mat-1999 and Con-mat-2003, suggesting that nDFA returns more accurate results on community detection than DFA. For CoauthorshipsNet1589, DFA puts 1585 among 1589 nodes into one community, and nDFA puts 1025 among 1589 nodes into one community. Recall that $Q(nDFA)$ is much larger than $Q(DFA)$ for CoauthorshipsNet1589, we see that DFA performs poor by tending to put nodes into one community while nDFA performs nice for returning a reasonable community structure. For CoauthorshipsNet379, though the overall embeddedness of nDFA is larger than DFA, nDFA's $\tau$ is much smaller than DFA, which suggests that nDFA returns more reasonable community partition for CoauthorshipsNet379 than DFA since DFA puts almost all nodes into one community. For Con-mat-1999 and Con-mat-2003, though $OE(DFA)$ is larger than $OE(nDFA)$, DFA again puts almost all nodes into one community for its large $\tau$. For run-time, we see that nDFA processes real-world weighted networks of up to 28000 nodes within tens of seconds. Generally, we see that $nDFA$ returns larger general modularity, smaller $\tau$ than that of DFA, suggesting nDFA provides more reasonable community partition. For comparative evaluation, simply using the overall embeddedness $OE$ is not enough, and we should combine $OE$ and $\tau$ for comparative analysis. Method returns larger $OE$ and smaller $\tau$ returns more reasonable community division, while method with larger general modularity always enjoys larger $OE$ and smaller $\tau$, i.e., $Q$ functions similar as larger $OE$ and smaller $\tau$ when a method gives reasonable community partition, just as how our nDFA performs on all real-world networks used in this paper. And this supports the effectiveness of our general modularity.
\section*{Conclusion}
In this paper, we introduced the Degree-Corrected Distribution-Free Model (DCDFM), a model for community detection on weighted networks. The proposed model is an extension of previous Distribution-Free Models by incorporating node heterogeneity to model real-world weighted networks in which nodes degrees vary, and it also extends the classical degree-corrected stochastic blockmodels to weighted networks by allowing connectivity matrix to have negative elements and allowing elements of adjacency matrix $A$ generated from arbitrary distribution as long as the expectation adjacency matrix $\Omega$ enjoys the block structure in Eq (\ref{DefinMM}). We develop an efficient spectral algorithm for estimating nodes labels under DCDFM by applying k-means algorithm on all rows in the normalized eigenvectors of the adjacency matrix. Theoretical results obtained by delicate spectral analysis guarantee that the algorithm is asymptotically consistent. The distribution-free property of our model allows that we can analyze the behaviors of our algorithm when $\mathcal{F}$ is set as different distributions. When DCDFM degenerates to DFM or DCSBM, our theoretical results match those under DFM or DCSBM. Numerical results of both simulated and empirical weighted networks demonstrate the advantage of our method designed by considering the effect of node heterogeneities. Meanwhile, to compare performances of different methods on weighted networks with unknown information on nodes communities, we proposed the general modularity as an extension of Newman's modularity. Results of simulated weighted networks and real-world un-weighted networks suggest the effectiveness of the general modularity. The tools developed in this paper can be widely applied to study the latent structural information of both weighted networks and un-weighted networks. Another benefit of DCDFM is the potential for simulating weighted networks under different distributions. Furthermore, there are many dimensions where we can extend our current work. For example, $K$ is assumed to be known in this paper. However, for most real-world weighted networks, $K$ is unknown. Thus, estimating $K$ is an interesting topic. Some possible techniques applied to estimate $K$ can be found in \cite{newman2016estimating,saldana2017how,chen2018network}. Similar as \cite{cai2015robust}, studying the influence of outlier nodes theoretically for weighted networks is an interesting problem. Developing method for weighted network's community detection problem based on modularity maximization under DCDFM similar as studied in \cite{chen2018convexified} is also interesting. Meanwhile, spectral algorithms accelerated by the ideas of random-projection and random-sampling developed in \cite{zhang2022randomized} can be applied to handle with large-scale networks, and we can take the advantage of the random-projection and random-sampling ideas directly to weighted network community detection under DCDFM. We leave studies of these problems for our future work.
\section*{Data availability}
All data and codes that support the findings of this study are available from the corresponding author upon reasonable
request.
\bibliography{refDCDFM}
\section*{Acknowledgements}
This work was supported by the High level personal project of Jiangsu Province (JSSCBS20211218).
\section*{Author contributions}
Huan Qing is the sole author of the paper.
\section*{Competing interests}
The author declares no competing interests.
\section*{Additional information}
\textbf{Correspondence} and requests for materials should be addressed to Huan Qing.\\
\textbf{Publisher’s note} Springer Nature remains neutral with regard to jurisdictional claims in published maps and
institutional affiliations.
\end{document}